\documentclass[12pt,a4paper]{article}
\usepackage[utf8]{inputenc}
\usepackage[T1]{fontenc}
\usepackage{amsmath,hyperref,bbm,cite}
\usepackage{amsfonts}
\usepackage{amssymb,comment}
\usepackage{graphicx}
\usepackage{notoccite}
\usepackage[left=2cm, right=2cm]{geometry}
\usepackage{todonotes}
\usepackage{xcolor}
\usepackage{physics}
\usepackage{bbold}
\usepackage{ntheorem}
\usepackage[overlay,absolute]{textpos}

\newcommand{\be}{\begin{equation}}
\newcommand{\ee}{\end{equation}}
\newcommand{\bea}{\begin{eqnarray}}
\newcommand{\eea}{\end{eqnarray}}

\newcommand\PlaceText[3]{%
\begin{textblock*}{10in}(#1,#2)
#3
\end{textblock*}
}%

\begin{document}

\PlaceText{165mm}{15mm}{19 May 2022}
\begin{center}
{\LARGE\bf A Local Wheeler-DeWitt Measure}

\vspace*{.5cm}
{\LARGE\bf for the String Landscape}

\vspace{.5cm}
\end{center} 

\vspace*{0.8cm}
\thispagestyle{empty}

\centerline{\large 
Bjoern Friedrich$^{1,}$, Arthur Hebecker$^1$, Manfred Salmhofer$^1$}
\vspace*{.2cm}
\centerline{\large Jonah Cedric Strau\ss$^1$, Johannes Walcher$^{1,2}$
}
\vspace{0.5cm}
 
\begin{center}
$^1${\it Institute for Theoretical Physics, Heidelberg University,}
\\
{\it Philosophenweg 19, 69120 Heidelberg, Germany}\\[0.5ex]  
\end{center} 
\vspace{-.4cm}
 
\begin{center}
$^2${\it Mathematisches Institut der Universit\"at Heidelberg,}
\\
{\it Im Neuenheimer Feld 205, 69120 Heidelberg, Germany}\\[0.5ex]  
\end{center} 
\vspace{.25cm}
\centerline{\small\textit{E-Mail:} \href{mailto:friedrich@thphys.uni-heidelberg.de}{friedrich@thphys.uni-heidelberg.de}, \href{mailto:a.hebecker@thphys.uni-heidelberg.de}{a.hebecker@thphys.uni-heidelberg.de},}
\centerline{\href{mailto:salmhofer@uni-heidelberg.de}{salmhofer@uni-heidelberg.de}, \href{mailto:j.strauss@thphys.uni-heidelberg.de}{j.strauss@thphys.uni-heidelberg.de},}
\centerline{\href{mailto:walcher@uni-heidelberg.de}{walcher@uni-heidelberg.de}}

\vspace*{.5cm}
\begin{abstract}\normalsize

\vspace*{.4cm}
\noindent
According to the `Cosmological Central Dogma', de Sitter space can be viewed as a quantum mechanical system with a finite number of degrees of freedom, set by the horizon area. We use this assumption together with the Wheeler-DeWitt (WDW) equation to approach the measure problem of eternal inflation. Thus, our goal is to find a time-independent wave function of the universe on a total Hilbert space defined as the direct sum of a variety of subspaces: A finite-dimensional subspace for each de Sitter vacuum and an infinite-dimensional subspace for each terminal Minkowski or AdS vaccuum. We argue that, to be consistent with semiclassical intuition, such a solution requires the presence of sources. These are implemented as an inhomogenous term in the WDW equation, induced by the Hartle-Hawking no-boundary or the Linde/Vilenkin tunneling proposal. Taken together, these steps unambiguously lead to what we would like to think of as a `Local WDW measure,' where `local' refers to the fact that the dS part of the resulting wave function describes a superposition of static patches. The global 3-sphere spatial section of the entire multiverse makes no appearance.
\end{abstract}
\vspace{10pt}

\underline{Keywords:} String Landscape, Multiverse, Wheeler-DeWitt equation, Measure Problem

\newpage

\tableofcontents 

\vspace*{1cm}

\section{Introduction}
In theories allowing for eternal inflation \cite{Steinhardt:1982kg, Vilenkin:1983xq, Guth:2007ng}, the measure problem \cite{Linde:1993nz} represents a fundamental challenge on the way to making statistical predictions in a multiverse. The reason is that, in such settings, `everything that can happen will happen infinitely many times,' such that the naive approach of counting becomes insufficient to define a probability measure. Of course,
the question about probabilities also arises in fundamental theories with multiple vacua which get populated in some other way. While the measure problem has gained popularity after the advent of the string theory landscape \cite{Bousso:2000xa, Kachru:2003aw, Susskind:2003kw, Denef:2004ze} (see \cite{Denef:2004ze, Schellekens:2015zua, Hebecker:2020aqr} for reviews), it is presumably a general and fundamental question of quantum gravity, possibly independent of its UV completion. Moreover, the measure problem does not disappear if, as recently conjectured, quantum gravity does not allow for metastable de Sitter vacua or for eternal inflation \cite{Danielsson:2018ztv, Ooguri:2018wrx, Bedroya:2019snp} (see \cite{Bedroya:2019tba, Rudelius:2019cfh, Wang:2019eym, Blanco-Pillado:2019tdf, Brahma:2020cpy} for related work). Indeed, while the traditional landscape of de Sitter vacua would in this case disappear, the measure problem still arises as soon as two solutions with temporal cosmological acceleration, as observed in our universe, exist.
While our analysis is phrased in terms of a landscape of (metastable) de Sitter vacua, it will turn out that the resulting local Wheeler-DeWitt measure is also applicable to a multiverse which only contains vacua with a dynamical, decaying dark energy.

The simplest and most studied approach to the measure problem, the introduction of a late-time cutoff, appears to come unavoidably with a certain amount of arbitrariness (see \cite{Freivogel:2011eg, Vilenkin:2006xv} for reviews and \cite{Linde:1993nz, Linde:1993xx, Garriga:2005av, Bousso:2006ev, Susskind:2007pv, DeSimone:2008bq, Bousso:2008hz, Garriga:2008ks, Harlow:2011az} for a selection of specific proposals). We feel that it is more promising to approach the issue in quantum cosmology \cite{Vilenkin:1994ua, Vilenkin:1995nb, Mersini-Houghton:2006phg, nomura2011physical, Hertog:2011ky, Hartle:2016tpo}, although it is certainly not obvious that an unambiguous answer can be obtained with our present, limited understanding of quantum gravity. For recent work in both directions see e.g.~\cite{Carifio:2017nyb, Chiang:2018dju, Finn:2018krt, Jain:2019gsq, Khoury:2019yoo, Vilenkin:2019mwc, Khoury:2019ajl, Olum:2021pux, Khoury:2021grg, Khoury:2022ish}.

The main paradigm we adopt is that only physics inside the horizon is real and the ever-growing spatial $S^3$ of global de Sitter (dS) space is irrelevant \cite{Dyson:2002pf, garriga2013watchers, nomura2011physical, nomura2012static,Hartle:2016tpo}. This has been called `single-observer picture' since the physics inside the horizon may be viewed as belonging to a single observer, located at the center. While it is advantageous and maybe necessary to put the observer center stage in conventional quantum mechanics, we find this problematic in the present case because one of the questions one wants to ask is precisely about the probability of finding observers, especially observers emerging as a result of an inflationary/reheating period. One then has to introduce some additional, {\it abstract} observers to count the former. To us, such an additional level of observers appears to be an ad-hoc input, similar to the choice of a cutoff prescription.

To avoid such issues, we prefer to formalize the local approach in an observer-independent way: We base our analysis on the cosmological central dogma \cite{banks2000cosmological,susskind2021impossible} which states that, from the perspective of the static patch, de Sitter space is a quantum system with a finite-dimensional Hilbert space. In a landscape with several de Sitter vacua, we take this to imply that our total Hilbert space is constructed as the direct sum of such finite-dimensional Hilbert spaces. Each of these subspaces describes a static patch of one of the dS vacua. Semiclassically, transitions between these patches arise since, in any one of them, a bubble of a different dS vacuum can nucleate and grow until the static patch is completely filled out by the new vacuum. In our effective quantum-mechanical model, this is encoded in elements of the Hamiltonian which violate the block-diagonal structure coming with the decomposition of the Hilbert space just described.\footnote{An alternative interpretation of the cosmological central dogma is that the entire Hilbert space of the dS part of the landscape belongs to the highest-entropy metastable dS vacuum with lower-entropy dS spaces emerging as excitations of the former.
The dimensions of their Hilbert spaces are exponentially smaller such that, at our level of precision, it is irrevant whether we view them as subspaces of the Hilbert space of the highest-entropy dS or as extra spaces, to be added in a direct sum.} 

It has been argued early on \cite{Dyson:2002pf} that static-patch based approaches generically suffer from a Boltzmann brain problem.
This is avoided by including terminal vacua, such that the decay rates of all dS vacua are larger than the production rate of Boltzmann brains, see e.g. \cite{nomura2011physical,Freivogel:2011eg}. We also note that, due to the enormity of the string landscape, finding an anthropically suitable vacuum is computationally complex \cite{Denef:2006ad, Denef:2017cxt}, which might embank the spread of the wavefunction through the landscape, even under eternal inflation. This is reminiscent of Anderson localization, which has been discussed in the present context in \cite{Mersini-Houghton:2006phg, Mersini-Houghton:2014yoa, MersiniHoughton:2005im} and \cite{Podolsky:2007vg}.

Our next key ingredient is the Wheeler-DeWitt (WDW) \cite{DeWitt:1967yk} equation, which should clearly be used instead of ordinary quantum-mechanical time evolution. The WDW operator acting on the total Hilbert space still encodes tunnelling between vacua. As a result, generic solutions are wave-functions with non-vanishing projections on all subspaces.

In the WDW approach, the conventional Schr\"odinger time evolution of quantum subsystems arises through correlations with some semiclassical observable \cite{DeWitt:1967yk,Lapchinsky:1979fd,Banks:1984cw}. The latter is usually taken to be the scale factor of the global de Sitter 3-sphere. One may try to save this intuitive definition of time in our local approach by imagining some `time variable behind the horizon,' rooted in the huge horizon Hilbert space. We identify a number of conceptual problems related to this perspective and opt for a different, more fundamental approach: Time can arise locally due to correlations between different observables, one of which is semiclassical but does not need to be related to the expansion of de Sitter \cite{DeWitt:1967yk,Banks:1984cw, Rovelli:2009ee}. The structure of the Hilbert space and the requirement of a time-independent state of the universe are not affected by adopting this local perspective on time.

Due to the Boltzmann brain problem and since the string theory landscape certainly contains supersymmetric Minkowski and anti-de Sitter (AdS) solutions, the framework described above must be enriched by terminal vacua. Such an addition changes the features of a landscape: for example, the probabilities with which dS vacua in the landscape are seen by observers depend on decay rates to terminal vacua \cite{Garriga:2005av, Schwartz-Perlov:2006swo}. A key potential issue for us is that, semiclassically, tunneling from dS to terminal vacua is possible, but not the reverse process. This can only be implemented by a stationary solution $\Psi$ if the corresponding probability current has a source in the dS part of the Hilbert space and `runs off to infinity' in the infinite-dimensional subspaces belonging to the terminal vacua. The source describes the creation of dS vacua out of nothing according to one of the proposals of \cite{hartle1983wave, Linde:1983mx, Vilenkin:1984wp}. At the fundamental level, this is implemented by adding an inhomogeneous term $\chi$ to the WDW equation:
\be
H\Psi=\chi\,.\label{iwdw}
\ee
We will not be able to specify $H$ and $\chi$ sufficiently well to solve for $\Psi$. Instead, we will argue that \eqref{iwdw} effectively leads to statements about probability currents and eventually to a simple rate equation for the probability distribution on the vacua of the multiverse. We solve this equation for a simple toy model, illustrating that the distribution is in some cases mainly determined by the amplitude for vacuum creation and the decay rates to terminal vacua. Tunneling between dS vacua could play a subdominant role. A detailed study of possible physics implications is left for future work.

The structure of the paper is as follows:
In Sect.~\ref{nvi} we develop a naive model for eternal inflation that is based on the cosmological central dogma and standard quantum mechanical time evolution. We show that semiclassical expectations are consistent with an appropriate generalization of Shnirelman's theorem.
In Sect.~\ref{sect_time_behind_the_horizon} we turn to the more appropriate WDW formalism, maintaining the conventional identification of time with the growing scale factor of de Sitter space.
We discover that this approach has various conceptional issues, prompting us to adopt a local perspective on time in Sect.~\ref{sect_Banks_approach}:
Here, no universal time variable is introduced. Instead time emerges independently for every experiment, through correlations with appropriate semiclassical observables.
In Sect. \ref{sect_terminal_vacua}, we extend the framework just developed by allowing for terminal vacua. Since the wave function of the universe must remain stationary, this entails either tunneling processes from terminals to dS or the presence of sources. Guided by semiclassics, we choose the latter option, i.e.~we allow for the creation of dS vacua from nothing. The resulting rate equation for the probability distribution is solved in Sect.~\ref{sect_applications} for a simple toy model containing two dS and one terminal vacuum. We briefly discuss how one could proceed from here to derive predictions relevant for post-inflationary observers like ourselves but leave a serious study of this subject to future work.

\section{The naive, quantum-mechanical perspective}\label{nvi}
\subsection{Density of states}
We embrace the notion of de Sitter space as a quantum mechanical system with a finite-dimensional Hilbert space~\cite{banks2000cosmological, Banks:1984cw, Frolov:2002va,Chandrasekaran:2022cip}. Based on this, we want to develop a simple toy model for eternal inflation, which is most naively understood as a sequence of tunneling transitions between several such de Sitter vacua.\footnote{
As 
will be discussed later on, the Schr\"odinger equation should in fact be replaced by the Wheeler-DeWitt (WDW) equation. For the moment we assume that (in some approximation) we can instead use a conventional Hamiltonian and a time variable defined e.g.~by some physical system (a clock)~\cite{Banks:1984cw, banks2000cosmological, Rovelli:1990jm, Rovelli:2009ee}. A more detailed analysis will be given in Sections \ref{sect_time_behind_the_horizon} and \ref{sect_Banks_approach}.
} 
The Hilbert space $\cal H$ of a single dS vacuum is of dimension $N\simeq\exp(S)$, where $S\simeq\pi(RM_P)^2$ denotes the entropy of the horizon.
We expect the energy eigenvalues of our dS space to form a very dense discretuum, at least in the energy range that corresponds to what we view as semiclassical de Sitter space. It then makes sense to introduce the continuous function $N(E)$ characterizing the number of states below an energy $E$ asymptotically, in a distributional sense.
We also introduce the density of states $\rho(E)=dN(E)/dE$.
Since there are only finitely many energy levels, $N$ and $N(E)$ clearly coincide for large enough $E$.
In this notation, the semiclassical dS states lie in a region of high density, presumably, their energies are close to $E_*$ with $E_*$ being defined as the value where $\rho(E)$ is maximal.
A small subsystem inside a dS space in a semiclassical state sees a temperature $T\sim 1/R$. 
The density of states at $E_*$ can then be computed using standard thermodynamic relations:
\be 
\frac{1}{R} \sim T=\left.\frac{dE(S)}{dS}\right|_{E_*}=\left.\frac{dE(S)}{d\log(N(E))}\right|_{E_*}=N(E_*)\left.\frac{dE(N)}{dN(E)}\right|_{E_*}=\frac{N(E_*)}{\rho(E_*)}\, .
\ee
Solving for $\rho(E_*)$ yields
\be
\rho(E_*)\sim RN(E_*)\, .
\ee
We naturally expect that $N (E_*)\sim N$. If we are willing to neglect sub-exponential terms, we may simply write
\begin{align}
    \rho(E_*)\sim N\, .\label{Density_of_states_estimate}
\end{align}
Note that this fixes the density of states in the energy range $E\sim E_*$ relevant for semiclassical dS, but it does not fix the precise value of $E_*$. The latter is not relevant in our context.

\subsection{Tunneling in the quantum mechanical toy model}\label{tqm}
Now, we allow for a second dS minimum, such that we can label the two dS vacua as \textit{false} and \textit{true} respectively. Our Hilbert space is then the direct sum of two subspaces of dimensions $N_f$ and $N_t$, where the index stands for true or false. We assume that $N_f\ll N_t$ and that tunneling transitions can occur. The total Hamiltonian takes the form
\be
H=\left(
\begin{array}{cc} H_f & \Delta \\ \Delta^\dagger & H_t \end{array}
\right)\,,\label{Hamilton}
\ee
where the $(N_f\times N_t)$-matrix $\Delta$ is assumed to have extremely small entries of typical size $\delta$.

We want to determine which $\delta$ matches the semiclassical expectation for the decay rate of a generic state in ${\cal H}_f$. In the field-theoretic regime, this rate is given by the Coleman-De Luccia (CDL) result \cite{coleman1980gravitational} $\Gamma_{sc}\simeq R_f^3\exp(-B)=R_f^3\exp(-S_B+S_A)$, with $S_B$ the action of the $O(4)$-symmetric bounce and $S_A$ the action of the false vacuum. The probability for a quantum state being in ${\cal H}_f$ then decays in the standard exponential way: \mbox{$p_t\sim\exp(-\Gamma_{sc} t)$}.\footnote{It
would be very interesting to also consider de Sitter minima which are connected by paths through flat scalar potentials, such that Hawking-Moss (HM) \cite{Hawking:1981fz} transitions become relevant. 
Also, for pairs of vacua that are not connected by an instanton, non-tunneling transitions \cite{Brown:2011ry} have to be considered.
We leave this subject to future work.}

\begin{figure}
	\centering
	\includegraphics[width=0.5\linewidth]{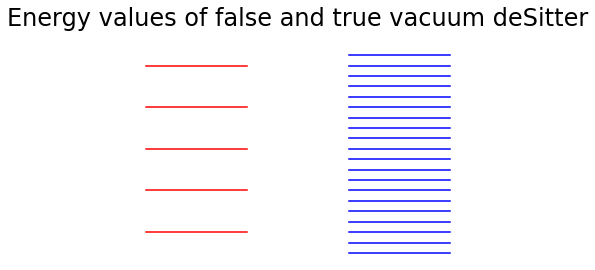}
	\caption{Energy spectra of the two dS vacua. Most eigenvalues of the false vacuum (left) lie inside the dense region of eigenvalues of the true vacuum (right).}%
	\label{fig_eigenvalue_distribution}
\end{figure}

To make progress, we now introduce a key assumption, illustrated in Fig.~\ref{fig_eigenvalue_distribution}:
We assume that the interval in which the energy eigenvalues of $H_f$ are mainly distributed lies inside the high-density region of the $H_t$ spectrum. As a result, the tunneling probability does not depend on which particular state of the false vacuum we are starting from. This agrees with semiclassical expectations since the large number of states comes from the horizon degrees of freedom, and the latter should not influence the semiclassical bubble formation probability of CDL.

Under the assumption above, $\rho_t$ is very large in the relevant energy range and a typical ${\cal H}_f$ state decays into a quasi-continuum. The rate in our quantum-mechanical toy model then follows Fermi's Golden Rule:
\be
\Gamma_{qm}\sim \delta^2\rho_t\,.\label{tunneling_rate_matrix_element}
\ee
Demanding $\Gamma_{qm}=\Gamma_{sc}$ allows us to estimate $\delta$ (disregarding non-exponential effects) as
\be
\delta^2 \sim \exp(-B)/\rho_t\,.\label{dsq}
\ee

We note that the standard derivation of Fermi's Golden Rule relies on a continuum approximation for the spectrum of the final-state subspace. This limits the validity to a time range below $T_{max}\sim 1/\rho_t$. If we want our false vacuum to actually decay within the time range we control, we need $\Gamma_{qm}\gtrsim 1/T_{max}$, which implies
\be
\rho_t\,\delta \gtrsim 1\,.\label{rdg}
\ee

The same constraint follows from a static perspective: Namely, we should demand that any eigenstate of the full system mostly points in the ${\cal H}_t$ part of the total Hilbert space. If this were not the case, one could prepare the system in an eigenstate pointing mostly in ${\cal H}_f$ and forever observe it in the false vacuum. This contradicts the semiclassical expectation that the false vacuum eventually has to decay. A necessary condition for our claim is that perturbation theory in $\delta$, correcting the original eigenstates of $H_f$, should break down. At leading order, the square of the ${\cal H}_t$-projection of a corrected $H_f$-eigenvector is estimated by
\be
\sum_{E_t}\frac{\delta^2}{(E_f-E_t)^2}\,.\label{sum}
\ee
This sum is dominated by the term that contains the eigenvalue $E_t^0$ which lies closest to $E_f$. We naturally expect that $E_f-E_t^0\sim 1/\rho_t$ (for an accidental resonance $E_f=E_t^0$, the cutoff at $T_{max}$ prevents a pole) such that the sum \eqref{sum} may be approximated by $\delta^2 \rho_t^2$. Requiring this to be of order unity or larger is equivalent to \eqref{rdg}. 

In our specific case of CDL tunneling, we have $\delta\sim \exp(-B)$ and $\rho_t\sim \exp(S_t)$, such that \eqref{rdg} turns into the condition
\be
B\lesssim S_t\,.\label{condition}
\ee
In fact, a general argument \cite{Kachru:2003aw} shows that instantonic tunneling transitions between dS vacua always obey
\begin{align}
    B\leq S_{dS}\,, \label{KKLT_bound}
\end{align}
with $S_{dS}$ being the entropy of the decaying dS.  This is consistent with expectations that dS spaces do not live longer than their recurrence time $t_{rec}\simeq \exp(S_{dS})$ \cite{Dyson:2002pf,Goheer:2002vf}. 
Since $S_f\leq S_t$, it is immediately clear that \eqref{KKLT_bound} implies \eqref{condition}.
The fact that the relevant bound \eqref{condition} holds on such general grounds encourages us to think that our basic picture of finite-dimensional Hilbert spaces, linked by decays into a quasi-continuum, is sensible.

We note that certain decay rates may violate our bound, e.g. the non-tunneling transitions of \cite{Brown:2011ry}. This does not represent a problem for our approach since there is no reason to expect that dS spaces exist which decay {\it exclusively} via such extremely slow transitions.

\subsection{Ergodicity and distribution of eigenvectors}\label{ede}
We expect the overwhelming majority of eigenvectors to deviate only very little from the $\mathcal{H}_t$ subspace of the Hilbert space $\mathcal{H}_t\oplus \mathcal{H}_f$. To quantify this, we consider two general classes of models. First, we draw an analogy to classical mechanics:
Our full Hilbert space with a Hamiltonian of the form \eqref{Hamilton} may be compared to a classical system the phase space of which is the union of two distinct subspaces. Moreover, transitions between these subspaces are suppressed.
For example, one may think of a particle in a large container with a peculiar geometry: Namely, it should be the union of two smaller containers connected by a short and narrow throat.

Now, we know that a particle in an irregular, compact geometry in general represents an ergodic system. In other words, integrating over a long period of time the (appropriately averaged) distribution of the particle position on a fixed-energy phase space slice is constant.
Moreover, the Shnirelman theorem \cite{shnirel1974ergodic} extends this statement to quantum mechanics. Specifically, it states that for a manifold with ergodic geodesic flow, the probability density associated to an eigenfunction of the Laplacian will generically, in the limit of large eigenvalues, tend towards the uniform distribution in phase space.

In principle, the Shnirelman theorem applies also to our geometry of interest, where the configuration space manifold consists of two well-separated regions connected by a narrow throat. However, in practice one is of course not interested in asymptotically high-energy eigenfunctions but rather hopes that the statement of the theorem holds with sufficient precision above a certain moderate energy threshold. The problem in our case is that this energy threshold presumably grows as the throat becomes more and more narrow. This narrowness is related to the smallness of $\delta$ in \eqref{Hamilton}. While it would be interesting to investigate the relevant limiting procedures more carefully, this would in itself not suffice to establish the applicability of Shnirelman's theorem. 
The reason is that there is another even more obvious obstacle: The Shnirelmann theorem, as it stands, only applies to infinite-dimensional Hilbert spaces where the Hamiltonian is an elliptic operator, for example the Laplacian.
Our system of interest has by definition a finite Hilbert space to start with. Moreover, we do not know whether it can be derived in some way from a related classical system.

Thus, the best we can do for now is to take the Shnirelman theorem as an inspiration and to proceed from the expectation that an analogous statement holds for our model Hamiltonian \eqref{Hamilton}. Our main motivations are the very large dimensions of the relevant subspaces and the expected random behaviours of the spectra of $H_{f,t}$ as well of the entries of the transition matrix $\Delta$.
To formlate our expectation quantitatively, we denote the norm of the projection of a vector on the $\mathcal{H}_{f,t}$ part of the total Hilbert space by $\Vert\cdot\Vert_{f,t}$. Now the claim that the probability density associated to an eigenvector tends towards the uniform distribution in phase space translates to the statement that, for a generic normalized eigenvector $v$ of $H$, one has
\be
\Vert v\Vert_f^2 \simeq \frac{N_f}{N_f+N_t}\,\, , \qquad\Vert v\Vert_t^2 \simeq \frac{N_t}{N_f+N_t} \,.\label{Shnirelman_ratio}
\ee
Here, the squared norms of the projections of $v$, i.e.~the left-hand sides in the two relations in \eqref{Shnirelman_ratio}, correspond to the probability distributions associated to $v$ in Shnirelman's sense. The right hand sides represent the relative sizes of the dimensions of the Hilbert subspaces, which corresponds to the claim of equidistribution in phase space volume.

The same result can be obtained using a second, totally different approach, based on different assumptions. To motivate it, recall that depending on the size of $\delta$ our system is found in two distinct regimes: For small $\delta$, perturbation theory in $\delta$ works and the $N_f+N_t$ eigenvectors split in two subsets -- $N_f$ of them lie mostly in ${\cal H}_f$, the other $N_t$ lie mostly in ${\cal H}_t$. The strong assumption we now make is that, when $\delta$ is not very small, the exact eigenvector basis is given by a randomly chosen orthonormal basis in ${\cal H}_t\oplus {\cal H}_f$.
We may specify this proposal by stating that our basis is the eigenbasis of a random hermitian matrix, as analysed in~\cite{benigni2021fluctuations,cipolloni2021normal}. The results of~\cite{benigni2021fluctuations,cipolloni2021normal} concerning the eigenvector distribution can be stated as follows: For a normalized eigenvector $v$ of a random hermitian matrix, the quantity $\Vert v\Vert_{f,t}^2$ is, in the large $N_f+N_t$ limit, Gaussian distributed with mean and variance 
\be
\mu_{f,t}=\frac{N_{f,t}}{N_f+N_t}\,\,\, ,\qquad \sigma^2_{f,t}\sim \frac{N_{f,t}}{(N_f+N_t)^2}\label{mean_var}\,.
\ee
In the limit of large $N_f$ and $N_t$, we see that the distribution of $\Vert v\Vert_{f,t}^2$ is sharply peaked around $\mu_{f,t}$. This is in perfect accordance with the proposal \eqref{Shnirelman_ratio} motivated using Shnirelman's theorem. Thus, whenever one de Sitter vacuum has an appreciably higher cosmological constant than the other, $N_t$ is exponentially larger than $N_f$, and hence $\Vert v\Vert_f^2$ will be exponentially small for a generic normalized eigenvector $v$. This conforms with the semiclassical expectation that the false vacuum unavoidably decays.

More specifically, we want to take the point of view of an observer in our landscape of two vacua and ask which vacuum they expect to find themselves in. In our approach, they can make a prediction by taking the solution of the Schr\"odinger equation describing their universe and projecting on one or the other vacuum. The most general solution is of the form
\begin{align}
\Psi = \sum_{i=1}^{N_t+N_f}a_i\exp(i\lambda_i t)v_i\,,\label{Schroedinger_solution}
\end{align}
where $\lambda_i$ are the eigenvalues of the full Hamiltonian and $v_i$ are the corresponding normalized eigenvectors. The complex coefficients $a_i$ parameterize the family of solutions.
The probability with which an observer sees the false vacuum is then given by
\begin{align}
R_{qm}=\frac{\int dt \Vert\Psi\Vert_f^2}{\int dt \Vert\Psi\Vert^2}=\frac{\sum_{i=1}^{N_f+N_t} \Vert v_i\Vert_f^2 |a_i|^2}{\sum_{i=1}^{N_f+N_t} |a_i|^2}
\simeq \frac{N_f}{N_f+N_t}\,.
\label{ratio_expr}
\end{align}
Here the last, approximate equality holds only for a generic solution or, more precisely, it is the expectation for an arbitrary but fixed choice of the $a_i$. 
Our quantum-mechanical result $R_{qm}$ may be compared with the semiclassical prediction based on following a single observer through the multiverse and integrating the times they spend in each vacuum \cite{garriga2013watchers, Vanchurin:2006qp}
\begin{align}
R_{sc}\sim\frac{\exp(-24\pi^2M_P^2/\Lambda_f)}{\exp(-24\pi^2M_P^2/\Lambda_f)+\exp(-24\pi^2M_P^2/\Lambda_t)}\sim \frac{N_f}{N_f+N_t}\, .
\label{rsc}
\end{align}
This result was derived from rate equations describing an infinite sequence of CDL tunneling events. It agrees with the prediction of our quantum mechanical toy model at the level of exponentially large terms. 

As noted in \cite{garriga2013watchers}, the non-exponential prefactors associated with the tunneling rates violate ergodicity, which in our approach is the basic starting point. At the technical level, this is not an immediate problem for us: On the one hand, we take the quantum mechanical picture more seriously than the semiclassical perspective of following a single observer's worldline. On the other hand, our present work merely attempts to outline an approach that works in principle -- we are happy to accept sub-exponential errors. Nevertheless, it would be interesting to pursue this apparent non-ergodicity issue in more detail. For example, our suggestion for a generalized Shnirelman theorem may have sub-exponential errors. It is also conceivable that the tunneling rate has corrections reinstating ergodicity (although we understand that the authors of \cite{garriga2013watchers} dismiss this option).

\section{Wheeler-DeWitt and `time behind the horizon'} \label{sect_time_behind_the_horizon}
This section represents our first attempt to take the crucial step from time-dependent quantum mechanics to the WDW approach. We will work with the usual assumption that time is related to the scale factor of the global dS sphere. This will lead to problems and will be abandoned in Sect.~\ref{sect_Banks_approach} in favor of a more radical approach, where no universal time variable is singled out.

In more detail, an outline for the present section reads as follows: We start by recalling the worldline action of a point-particle moving in target space $\{X^\mu\}$. Worldline reparametrization invariance gives rise to a WDW-type constraint equation, in which target-space time $X^0$ is singled out by its wrong-sign kinetic term. Hence, also in the worldline view,  the argument $X^0$ of the WDW wave function may naturally be interpreted as `time'. 
Moreover, this toy model serves us to discuss the options of employing the Klein-Gordon or the Hartle-Hawking scalar product \cite{Perry:2021udd}. While the former has conceptual advantages, one encounters problems when trying to predict measurements such that we eventually settle on the latter.

We then move on to a further toy model: The mini-superspace description of a homogeneous and isotropic universe. Here, the scale factor appears with a wrong-sign kinetic term
and may be taken to play the role of time. This model has been adopted in 
\cite{MersiniHoughton:2005im,Mersini-Houghton:2006phg, Mersini-Houghton:2014yoa} to derive a measure on the landscape from a WDW equation. However, since the nucleation of pocket universes breaks spacetime homogeneity, we find such a global mini-superspace approach unsatisfying.
We then make an attempt to unify the static-patch-based view of the multiverse with the intuition obtained from the point-particle and mini-superspace discussion.
We speculate that the expansion dynamics of the global dS three-sphere is, in the local approach, encoded in the degrees of freedom associated to the dS horizon.
We study an effective Hamiltonian appropriate for this kind of model but we quickly encounter various issues, related for example to ambiguities in the structure of the total Hilbert space. As a result, we find it 
more promising to completely abandon the scale factor as a dynamical variable and pursue, 
in Sect. \ref{sect_Banks_approach}, a strictly local approach where time emerges entirely within the horizon, independently of the expansion of global de Sitter space. 

The reader not interested in our attempts to develop the approach of 
\cite{MersiniHoughton:2005im,Mersini-Houghton:2006phg, Mersini-Houghton:2014yoa} may directly move on to Sect. \ref{sect_Banks_approach}.

\subsection{From particle worldlines to mini-superspace and landscape\texorpdfstring{\\}{} dynamics}\label{pwl}
We called the discussion of Sect.~\ref{nvi} `naive' for an obvious reason: We are dealing with a time reparameterization invariant theory and the Schr\"odinger equation is hence not the appropriate tool. Instead, one has to use the 
Wheeler-DeWitt (WDW) equation \cite{DeWitt:1967yk} (see \cite{Wiltshire:1995vk} for a review),
\be 
H{\Psi} = 0\,.\label{WDW}
\ee
While the wave function of the universe $\Psi$ is by definition time-independent, the concept of time naturally arises from correlations between observables.

We find it convenient to start with a particularly simple toy model - a point-particle moving in $d$-dimensional Minkowski space. Its Polyakov action reads
\begin{align}
S=\int d\tau \, \left(\gamma^{-1} \dot{X}^\mu (\tau )\dot{X}^\nu (\tau ) \eta_{\mu\nu}-\gamma\left(m^2+V\left(X^i (\tau )\right)\right)\right)\, ,\label{WDW_Lagrangian}
\end{align}
where $X^\mu$ are the target space coordinates, $\gamma(\tau)$ is the square root of the worldline metric (the einbein), and $V(X^i)$ is a potential in which the particle is trapped. For simplicity, $V$ does not depend on $X^0$. From the reparameterization invariance of $\tau$, one derives a constraint which, after quantization, takes the form
\be
    H\Psi(X^\mu)=0\qquad \mbox{with}\qquad 
    H=\frac{1}{2}(-\partial_\mu\partial^\mu +m^2 +V(X^i))\, .\label{Worldline_Hamiltonian}
\ee
From the WDW perspective, the coordinate $X^0$ is set apart since it enters the equation with a wrong-sign kinetic term. Since $V$ does not depend on $X^0$, a product ansatz of the form \mbox{$\Psi(X^\mu)=\exp(\pm ikX^0)\chi (X^i)$} can be made. This leads to a time-independent Schr\"odinger equation for $\chi$:
\begin{align}
    (-\partial_i\partial^i + m^2 +V(X_i))\chi(X^i)=k^2\chi(X^i)\, .
\end{align}
We see that, not only in target space but also in the resulting quantum mechanical system, $X^0$ plays the role of time. 
The only special feature is the appearance of positive and negative frequency modes -- a consequence of the WDW equation being second rather than first order in $\partial_0$.

On the space of wavefunctions $\Psi (X^\mu)$, a natural scalar product is given by
\begin{align}
    (\Psi,\Phi)=\int \mathrm{d}^dX\, \Psi^*\Phi\,.\label{Hartle_Hawking_product_particle}
\end{align}
However, the subspace of solutions to \eqref{Worldline_Hamiltonian} is in addition equipped with an alternative inner product. To see this note that \eqref{Worldline_Hamiltonian} is (a slightly modified version of) the Klein-Gordon equation, implying that the current
\begin{align}
j^\mu=\frac{i}{2}\left(\Psi\partial^\mu \Psi^* - \Psi^*\partial^\mu\Psi\right)\label{KG_current}  \end{align}
is conserved. This allows for the definition of the Klein-Gordon inner product
\begin{align}
    (\Psi,\Phi)_{KG}=\int_{\Sigma}  \frac{i}{2}\left(\Phi\partial_\mu \Psi^* - \Psi^*\partial_\mu \Phi\right) \mathrm{d}A^\mu\,.\label{KG_inner_product}
\end{align}
Crucially, assuming appropriate boundary conditions, this is independent of the choice of the spacelike codimension-one submanifold $\Sigma$ over which the current is integrated.
However, taking \eqref{KG_inner_product} as the Hilbert space inner product of the quantum theory has issues. The first one is the well-known fact that \eqref{KG_inner_product} is only positive definite if restricted to positive frequency solutions.
The second problem is related to calculating expectation values of operators $O$.
In general, $O\Psi$ does not satisfy \eqref{Worldline_Hamiltonian}, implying that a current analogous to \eqref{KG_current} but including $O$,
\begin{align}
j^\mu=\frac{i}{2}\left((O\Psi)\partial^\mu \Psi^* - \Psi^*\partial^\mu (O\Psi)\right)\,,  
\end{align}
is not conserved. As a result, $(\Psi,O\Psi)_{KG}$ depends on the chosen hypersurface $\Sigma$. To illustrate this, consider \eqref{Worldline_Hamiltonian} in $1+1$ dimensions with vanishing mass and potential. Then, a simple wave packet,
\begin{align}
    \Psi (X^0,X^1) =\exp(-(X^1-X^0)^2+ik_0(X^1-X^0))\,,\label{wave_packet}
\end{align}
represents a solution.
We should be able to predict the most likely value of $X^1$ when $X^0$ takes values between $X^0_a$ and $X^0_b$.
Put differently, we want to calculate the expectation value of $P^0_{ab}X^1$, where $P^0_{ab}$ is the operator projecting the wavefunction to the strip $X^0\in (X^0_a, X^0_b)$.
Using \eqref{Hartle_Hawking_product_particle}, we find the intuitive result
\begin{align}
    (\Psi,P^0_{ab}X^1\Psi)=\int_{X^0_a}^{X^0_b} \!\!dX^0\,\int_{-\infty}^\infty \!\!dX^1\, X^1\exp(-2(X^1-X^0)^2)=\frac{\sqrt{2\pi}}{4}(X^0_b-X^0_a)(X^0_b+X^0_a)\,.\label{Expectation_X1_Hartle_Hawking}
\end{align}
By contrast, the Klein-Gordon inner product with $\Sigma$ chosen to be the hypersurface $X^0=\,$const. gives \footnote{For $k_0\gg 1$, the wave packet \eqref{wave_packet} contains essentially only positive frequency modes, such that the issue of non-positive-definiteness is absent.}
\begin{align}
    (\Psi,P^0_{ab}X^1\Psi)_{KG}=\begin{cases}
        0 & \textbf{if } X^0\notin (X^0_a,X^0_b)\\
        \int dX^1\, k_0 X^1\exp(-2(X^1-X^0)^2)=\sqrt{\frac{\pi}{2}}k_0X^0 & \textbf{if } X^0\in (X^0_a,X^0_b)
    \end{cases}\,.\label{Expectation_X1_KG}
\end{align}
We observe that \eqref{Expectation_X1_KG} depends on the chosen value of $X^0$, which is expected since $P^0_{ab}X^1\Psi$ does not solve \eqref{Worldline_Hamiltonian}. Of course, to obtain a result independent of $X^0$ one may consider averaging \eqref{Expectation_X1_KG} over all hypersurfaces $X^0=\,$const. within the prescribed interval.
But this appears ad hoc and goes against the key idea that the Klein-Gordon inner product should be hypersurface-independent.

The problem of $(\Psi,O\Psi)_{KG}$ not being well defined occurs generally for operators that do not commute with $H$, which is the case for most operators of practical interest. This lends support to the use of \eqref{Hartle_Hawking_product_particle} rather than \eqref{KG_inner_product}.

A more relevant model for us is the mini-superspace description of an FLRW spherical geometry with scale factor $a=\exp(\alpha)$, curvature $k$, as well as a (homogeneous) scalar field $\phi$ in a potential $V(\phi)$. The appropriately truncated Einstein-Hilbert action reads \cite{DeWitt:1967yk, hartle1983wave}
\be
  S=\int\dd t\left[-e^{-3\alpha}\dot{\alpha}^2+ke^\alpha+e^{3\alpha}\dot{\phi}^2-e^{3\alpha}V(\phi)\right],
\ee
showing that this time the scale factor plays the special role of the variable with the wrong-sign kinetic term. The resulting WDW operator takes the form
\be
  H=\frac12\left[e^{-3\alpha}\left(\partial_\alpha^2+p'(\alpha)\partial_\alpha-\partial_\phi^2\right)+e^{3\alpha}\left(V(\phi)-ke^{-2\alpha}\right)\right]\, ,\label{MSSM_Hamilton}
\ee
where the function $p(\alpha)$ parameterizes an operator ordering ambiguity \cite{partouche2021wavefunction, hartle1983wave}:
\be
  e^{-p(\alpha)}\pi_\alpha e^{p(\alpha)}\pi_\alpha\quad\to\quad -e^{-p(\alpha)}\partial_\alpha\left( e^{p(\alpha)}\partial_\alpha\right)\,.
\ee
In the late-time, semiclassical regime (i.e.~$\alpha\to\infty$), the WDW equation simplifies to 
\be
  0=\left[e^{-3\alpha}\left(\partial_\alpha^2-\partial_\phi^2\right)+e^{3\alpha}V(\phi)\right]\Psi(\alpha, \phi)\,.\label{MSSM_WDW}
\ee
In analogy to \eqref{Hartle_Hawking_product_particle}, one may use the scalar product
\be
\left\langle\Psi_1\middle|\Psi_2\right\rangle = \int d\phi \int_{-\infty}^\infty\mathrm{d}\alpha \,e^{3\alpha}\,\Psi_1^*(\alpha,\phi)\Psi_2(\alpha,\phi)\,,\label{Hartle_Hawking_inner_product}
\ee
where the factor $e^{3\alpha}$ ensures that the WDW operator is self adjoint \cite{partouche2021wavefunction,hartle1983wave}.

Alternatively, one could have tried to use the DeWitt inner product \cite{DeWitt:1967yk} on the space of solutions to \eqref{MSSM_WDW}, which is defined in analogy to the Klein-Gordon inner product \eqref{KG_inner_product},
and has recently proven to be a useful tool in holographic computations \cite{Maldacena:2019cbz,Iliesiu:2020zld}.
As in the Klein-Gordon case, one faces the issue of non-positive-definiteness. Although a probability interpretation can nevertheless be given in special situations, e.g. when working in the WKB approximation where the wavefunction can be decomposed in positive and negative frequency parts, it is not clear how the DeWitt product can be used in more general setups \cite{Wiltshire:1995vk,Hawking:1985bk,vilenkin1989interpretation,Kehagias:2021wwr}.
Moreover, operators corresponding to observables of interest, such as the counting of galaxies at large $\alpha$, fail to commute with $H$.
Thus, as in the Klein-Gordon case, making their expectation values well-defined would require an ad-hoc average over hypersurfaces.

Our preferred inner product definition \eqref{Hartle_Hawking_inner_product}, which one may call the `Hartle-Hawking inner product'~\cite{Perry:2021udd}, can be applied before restricting to solutions of \eqref{MSSM_WDW} and before gauge fixing. One of its main known problems are divergences arising in the integration over the entire parameter space.
We give two arguments in its defense: First, even if the integral in \eqref{Hartle_Hawking_inner_product} is divergent, one may still be able to derive finite relative probabilities. 
Second, in a theory with a finite-dimensional Hilbert space, which is what we are after eventually, the problem of diverging integrals is presumably entirely absent.
Finally, we note that divergent integrals may also arise when using the DeWitt inner product to calculate expectation values, namely in the process of averaging over hypersurfaces.

By comparison with the point particle example above, it now appears very natural to think of $\alpha$ as a time variable \cite{Kiefer:1992cn, Kiefer:1987ft}. One may then encode a (toy-model) landscape in a set of minima of $V(\phi)$ and try to approach the measure problem in the resulting mini-superspace framework. Such an approach has in fact already been pursued in \cite{MersiniHoughton:2005im,Mersini-Houghton:2006phg, Mersini-Houghton:2014yoa}\footnote{We found the explicit solutions of toy-model WDW equations with concrete scalar-field potentials in \cite{Mersini-Houghton:2014yoa} more convincing than the general analysis of \cite{MersiniHoughton:2005im,Mersini-Houghton:2006phg}. However, as will be explained in the main text, we have conceptual issues with this framework independently of the technical implementation.} 
and \cite{Podolsky:2007vg}. To be precise, only \cite{MersiniHoughton:2005im,Mersini-Houghton:2006phg, Mersini-Houghton:2014yoa} refer directly to the WDW equation while \cite{Podolsky:2007vg} appears to be closer to what we called `naive quantum-mechanical' approach. Both references appeal to Anderson localization \cite{anderson1958absence} to make statements about the wave function of the universe.

We see conceptual issues with the straightforward use of the mini-superspace model based on $\alpha$ and $\phi$ advocated in \cite{MersiniHoughton:2005im,Mersini-Houghton:2006phg, Mersini-Houghton:2014yoa}. The problem is that Coleman-De Luccia instantons and the subsequently expanding bubbles violate global homogeneity of the spatial 3-sphere. But this was a key assumption underlying the model. Put differently, tunneling in the mini-superspace model corresponds to transitions where the whole spatial $S^3$ changes the vacuum state at once. But this is not what happens according to Coleman-De Luccia and it is inconsisent with causality at late times, when the 3-sphere is much larger than the horizon.

By contrast, Ref.~\cite{Podolsky:2007vg} focuses on a single Hubble patch of a dS space and analyzes the scalar-field dynamics resulting from the nonzero temperature of the dS. The starting point is the Langevin equation (derived in this context by Starobinsky \cite{starobinsky1988stochastic}). From this, an effective Schr\"odinger-type equation can be obtained, to which Anderson's findings are eventually applied. The problem with this approach is that the Langevin equation is only derived in a regime where the slow roll conditions hold, but the later analysis makes statements about the entire landscape (where we do not expect transitions between different minima to occur in the slow roll regime). Still, it may in principle be possible and worthwhile to compare our results with those of \cite{Podolsky:2007vg} in a restricted setting where both are applicable.

\subsection{The time variable as part of the horizon degrees of freedom}\label{tbh}
Let us now try to develop an alternative suggestion for a WDW equation in a landscape situation, avoiding the apparent shortcomings of earlier approaches just discussed. As already noted at the very beginning of Sect.~\ref{sect_time_behind_the_horizon}, our results will still not be satisfactory and we will eventually turn to an improved, more minimalist point of view in Sect.~\ref{sect_Banks_approach}.

Crucially, we can not use a mini-superspace modelling the whole spatial $S^3$ since we assume that only the spacetime inside the horizon is real. The wave function ${\Psi}$ of the universe hence describes a superposition of different vacua in a single static patch. In general, the static patch looks homogeneous and isotropic to the observer sitting at the center. This homogeneity is broken if a bubble of a new vacuum nucleates, but it is restored very quickly as the de Sitter expansion continues in the new vacuum. Thus, one may hope that a quantum description similar to the mini-superspace approach exists. To write down an appropriate Hamiltonian, we may take \eqref{MSSM_WDW} as an inspiration. However, there appears to be no justification for the relative exponential factors between $\partial_\phi^2$ and $V(\phi)$ since the local $\phi$-field dynamics inside the horizon is not modified in the course of the ongoing exponential expansion. Let us then, in a first attempt, drop the exponential factors altogether and consider the WDW Hamiltonian
\be 
H=\partial_\alpha^2+H_\phi\,,\label{alpha_Hamilton}
\ee
acting on the Hilbert space\footnote{We view \eqref{alpha_Hamilton} as some form of `continuum approximation' to the true Hamiltonian acting on the finite-dimensional space \eqref{alpha_Hilbert_space_structure}.}
\begin{align}
    \mathcal{H}_{dS}=\mathcal{H}_{bulk} \otimes\mathcal{H}_{horizon}\,,\qquad \quad(\,\,\mbox{with}\,\,\,\,\dim \mathcal{H}_{dS}=\exp(S_{dS})\,\,)\,,
    \label{alpha_Hilbert_space_structure}
\end{align}
of the static-patch observer. This tensor product ansatz is inspired, among others, by \cite{nomura2011physical, nomura2012static}. In the simplest mini-superspace model, $H_\phi=-\partial_\phi^2+V(\phi )$. More generally, the Hamiltonian $H_\phi$ should describe the dynamics of all scalar-field and gravitational fluctuations of the static patch.  These degrees of freedom are encoded in $\mathcal{H}_{bulk}$, which is hence subject to the action of $H_\phi$. It remains at this stage unclear to which extent and how $H_\phi$ also acts on $\mathcal{H}_{horizon}$. Some form of non-trivial action of $H_\phi$ on $\mathcal{H}_{horizon}$, which of course contains the majority of degrees of freedom of the entire system \cite{Cohen:1998zx}, is presumably required. 
The variable $\alpha$ is interpreted as living in $\mathcal{H}_{horizon}$ and encoding a remnant of the semiclassical expansion of the global dS sphere which, of course, we have abandoned. Hence, the operator $\partial_\alpha^2$ acts only on the space $\mathcal{H}_{horizon}$.

We note that the approximate factorization in bulk and horizon degrees of freedom assumed in \eqref{alpha_Hilbert_space_structure} relies on the limit of weak gravity, $1/M_P^2\to 0$. It is strongly broken if the bulk region contains so much energy that the size of the horizon shrinks significantly. This happens e.g.~for black hole states close to the Nariai limit.

Comparing the Hamiltonian \eqref{alpha_Hamilton} with that in \eqref{Worldline_Hamiltonian} and the subsequent analysis, it becomes apparent that $\alpha$ plays the role of a time variable. We may furthermore attempt to identify $H_\phi$ with the Hamiltonian \eqref{Hamilton} of our toy-model quantum mechanical description. But such an identification is not perfect since some of the degrees of freedom are already encoded in $\alpha$.

Unlike in \eqref{Schroedinger_solution}, solutions of the WDW equation are now composed of positive and negative frequency parts
\begin{align}
    \Psi (\alpha )= \sum_{i=1}^{N_t+N_f}\left(a_i\exp(i\sqrt{\lambda_i} \alpha )+b_i\exp(-i\sqrt{\lambda_i} \alpha)\right)v_i\,,\quad a_i,b_i\in\mathbb{C}\, .\label{alpha_wavefunction}
\end{align}
We would like to argue that, nevertheless, the crucial calculation \eqref{ratio_expr} remains essentially unchanged:
\begin{align}
    R_{qm}=\frac{\int d\alpha |\Psi|_f^2}{\int d\alpha |\Psi|^2}=\frac{\sum_{i=1}^{N_f+N_t} \Vert v_i\Vert_f^2 \left(|a_i|^2+|b_i|^2\right)}{\sum_{i=1}^{N_f+N_t} \left(|a_i|^2+|b_i|^2\right)}
\sim \frac{N_f}{N_f+N_t}\,,\label{ratio_alpha}
\end{align}
Concerning the first two equalities this is obvious. However, arguing for the approximate result $N_f/(N_f+N_t)$ after the last step is problematic. Originally, we arrived at this relation by generalizing Shnirelman's theorem  and assuming $\Vert v_i\Vert_f^2/\Vert v_i\Vert_t^2 \sim N_f/N_t$ for a generic eigenvector $v_i$ in ${\cal H}_t\oplus {\cal H}_f$. In our present setting we would have to generalize \eqref{alpha_Hilbert_space_structure} according to
\be
{\cal H}_{tot} = {\cal H}_t \oplus {\cal H}_f 
= \Big( \mathcal{H}_{bulk}^t \otimes \mathcal{H}_{horizon}^t \Big) \,\oplus\,
\Big( \mathcal{H}_{bulk}^f \otimes \mathcal{H}_{horizon}^f \Big)
\ee
to make this argument. 

The difficulty is now that our variable $\alpha$ somehow has to belong to the two horizon Hilbert spaces $\mathcal{H}_{horizon}^t$ and $\mathcal{H}_{horizon}^f$ together. Even worse, in a more realistic case there would be as many such Hilbert spaces as there are dS vacua and our single variable $\alpha$ has to be encoded in all of them at the same time. Even though this sounds maybe surprising, it is certainly conceivable.

Alternatively, we may try to abandon the idea of deriving the last relation in \eqref{ratio_alpha} using Shnirelman's theorem. Then, we can speculate that the Hamiltonian $H_\phi$ only acts on states in $\mathcal{H}_{bulk}$. The action of \eqref{alpha_Hamilton} then factorizes according to the product structure  $\mathcal{H}_{bulk} \otimes \mathcal{H}_{horizon}$ of our full Hilbert space.
While the Shnirelman theorem now can not be used to argue for the result \eqref{ratio_alpha}, we may appeal to the more conventional approach of looking for stationary solutions of tunneling rate equations \cite{garriga2013watchers, Vanchurin:2006qp}. 
As discussed around \eqref{rsc}, this gives the same result: $R_{sc} \sim R_{qm} \sim N_f/(N_f+N_t)$. In this case, the appropriate generalization of \eqref{alpha_Hilbert_space_structure} would be
\be
{\cal H}_{tot} = \Big({\cal H}_{bulk}^t \oplus {\cal H}_{bulk}^f \Big) \otimes {\cal H}_{horizon}\,.
\ee
However, it then remains unclear how exactly to think about ${\cal H}_{horizon}$. After all, its dimension depends on the cosmological constant associated with the bulk, the latter in general being very different between true and false vacuum. 

Furthermore, we note that following \cite{nomura2012static} one might also imagine that additional degrees of freedom are present behind the horizon, encoded in a Hilbert space $\mathcal{H}_*$. In this case,  \eqref{alpha_Hilbert_space_structure} has to be replaced by
\be
{\cal H}_{dS} = {\cal H}_{bulk} \otimes {\cal H}_{horizon} \otimes {\cal H}_{*}\,, 
\label{tfa}
\ee
with\footnote{
Reference 
\cite{nomura2012static} calls our ${\cal H}_{bulk} \otimes {\cal H}_{horizon}$ the `bulk' and our ${\cal H}_{*}$ the `horizon' Hilbert space.
} 
$\mbox{dim}(\mathcal{H}_*)=\mbox{dim}(\mathcal{H}_{bulk}\otimes\mathcal{H}_{horizon})$.
This may be motivated by related proposals in the black hole context \cite{Nomura:2013nya, Verlinde:2013uja, Nomura:2013gna}. There, it is suggested that the smooth horizon of area $A$ is described by an entanglement subspace of dimension $\exp(A/4l_P^2)$ in a larger space of dimension $\exp(A/4l_P^2)\times \exp(A/4l_P^2)$. The last two factors characterize the exterior (near-horizon) and interior black hole degrees of freedom. The generalization to dS, with the obvious change of meaning of `internal' and `external,' has been discussed in \cite{Nomura:2013nya, Nomura:2013gna}. While in this way one might have found, in the form of ${\cal H}_{*}$, a place for the time variable $\alpha$ to live, we feel that this somehow morally clashes with our starting point of giving top priority to the cosmological central dogma. Moreover, while it then appears natural to generalize \eqref{tfa} to the (2-vacuum) landscape according to 
\be
{\cal H}_{tot} = 
\bigg( \Big( \mathcal{H}_{bulk}^t \otimes \mathcal{H}_{horizon}^t \Big) \,\oplus\,
\Big( \mathcal{H}_{bulk}^f \otimes \mathcal{H}_{horizon}^f \Big) \bigg) \otimes {\cal H}_{*}\,,
\ee
we are not certain that we have enough motivation for such a proposal.

Let us postpone any further discussion of how to derive the last step in \eqref{ratio_alpha} and turn to a different, potentially also problematic aspect of our quantum mechanical analysis. Namely, it may appear to be a radical step to simply drop the exponential factors when going from \eqref{MSSM_WDW} to \eqref{alpha_Hamilton}. We would like to argue that it is possible to proceed in a slightly more general manner.
The exponential factors appearing in front of $-\partial_\phi^2$ and $V(\phi)$ in \eqref{MSSM_WDW} have their origin in the expansion of global space. 
Since we are only considering the local theory inside a static patch of finite volume that does not change its size, there should be no exponential factors whatsoever multiplying the Hamiltonian $H_\phi$.
This can be verified by looking at a particle living in a de Sitter space which is localized at a specific point on the spatial $S^3$. 
A straightforward analysis of the equations confirms our intuition that there are no exponential factors appearing in front of the particle Hamiltonian in the resulting WDW equation.
Nevertheless, the factor of $e^{-3\alpha}$ multiplying the term $\partial_\alpha^2$ is still present.
Thus, a more appropriate form of the WDW equation taking into account the local dynamics of global expansion is given by
\be\label{wdw:simple-schroed}
  0=\left(e^{-3\alpha}\partial_\alpha^2+H_\phi\right)\Psi(\alpha).
\ee
After diagonalization of $H_\phi$, equation \eqref{wdw:simple-schroed} reduces to a number of Schr\"odinger-type equations, one for each eigenvalue $\lambda_i$ of $H_\phi$:
\be
  0=\left(\partial_\alpha^2+\lambda_ie^{3\alpha}\right)\psi_i(\alpha )\, .\label{Bessel_equation}
\ee
Upon substituting $e^{3\alpha}=x^2$ this turns into Bessel's differential equation and can hence be solved exactly in terms of Bessel functions. However, we only need the solutions at large $\alpha$, which can easily be obtained using quantum mechanical intuition. Indeed, \eqref{Bessel_equation} describes a particle with zero energy rolling down a potential $V(\alpha)=-\lambda_i\exp(3\alpha)$. This suggests the ansatz $A(\alpha)\, {\rm exp}(\pm iS(\alpha))$ with $S'=\sqrt{-V}$. The prefactor is fixed by noting that the semiclassical probability flux, which is constant along the half-line $\alpha>0$, may be written as the product of the probability density $\sim |A|^2$ and the velocity $v$. Hence, $|A|^2 \sim 1/v\sim 1/\sqrt{-V}$. In complete analogy to \eqref{Schroedinger_solution} and \eqref{alpha_wavefunction} we then have
\be
  \Psi(\alpha)=\sum_{i=1}^{N_f+N_t}e^{-3\alpha/4}\left( a_i \exp\left(\frac{2i}{3}\sqrt{\lambda_i}e^{3\alpha/2}\right)+b_i \exp\left(-\frac{2i}{3}\sqrt{\lambda_i}e^{3\alpha/2}\right)\right)v_i\, ,\quad a_i,b_i\in\mathbb{C}, 
\ee
where the $v_i$ are the eigenvectors of $H_\phi$ corresponding to $\lambda_i$.
To demonstrate the claim that the reintroduction of exponential factors in this form leaves physical predictions unchanged, we have to again look at expectation values such as \eqref{ratio_alpha}. Here, we encounter formal integrals 
\be
  \int_{-\infty}^\infty d\alpha\, e^{3\alpha}\,
  \left(e^{-3\alpha/4} \exp\left(\frac{2i}{3}\sqrt{\lambda_i}e^{3\alpha/2}\right)\right)\,
  \left(e^{-3\alpha/4} \exp\left(\pm\frac{2i}{3}\sqrt{\lambda_i}e^{3\alpha/2}\right)\right)\,.
  \label{Bessel_integral}
\ee
The factor $e^{3\alpha}$ is part of the definition of the integration measure which is chosen in order for the Hamiltonian \eqref{wdw:simple-schroed} to be hermitian, as discussed before.
Since the integral \eqref{Bessel_integral} is clearly dominated by the region of large $\alpha$, our restriction to large-$\alpha$ solutions is justified. Changing the integration variable to $x = e^{3\alpha/2}$ shows that the integral may be interpreted in the distributional sense as an oscillatory integral. Due to the Hilbert space being finite-dimensional, this integral should possess some built-in cutoff at large but finite $\alpha$. This may be modelled by introducing a smooth regulator at $\alpha\to\infty$. As a result, only terms $\sim |a_i|^2$ and $\sim |b_i|^2$ will survive. This reproduces the result of \eqref{ratio_alpha}.
Thus, the Hamiltonian \eqref{wdw:simple-schroed} generates the same predictions as already discussed around \eqref{ratio_alpha}.
We conclude that an appropriate inclusion of exponential factors originally coming from the dS expansion leads to the same phenomenological predictions as the naive calculation \eqref{ratio_alpha}.

Let us draw preliminary conclusions from this subsection: We tried to turn the scale factor variable $\alpha$ into some form of `time behind the horizon'. While this perspective follows rather naturally from trying to merge Wheeler-DeWitt and cosmological central dogma, it also has obvious troubles. In particular, we find it hard to decide whether there should be a separate horizon Hilbert space for each vacuum, or one for all vacua, or whether some compromise must be found. In the first case, which appears to be most consistent with our earlier discussion in Sect.~\ref{nvi}, time would have to be encoded by several Hilbert spaces collectively, in a way we do not understand. More generally, since we probably have to give up the eternally expanding spatial sphere of a global de Sitter space, we do not know how the variable $\alpha$ is microscopically related to the degrees of freedom behind the horizon. Is it possible that complexity, which in the black hole case is known to continue growing long after the black hole has settled to a stationary state, has a relation to our desired time variable? Finally, we are forced to accept the puzzling conclusion that time is {\it not} truly, mathematically non-compact. The reason is that a properly non-compact time could not be encoded in a Hilbert space of finite (no matter how large) dimension. Of course, an effective time variable running over a very large interval is presumably sufficient for all practical purposes. 

Given all these difficulties, we find it justified to pursue a more minimalist approach in the next section, abandoning any distinguished classical variable which plays the role of time.

\section{The perspective of a `WDW purist'}\label{sect_Banks_approach}
In \cite{DeWitt:1967yk,Banks:1984cw} it was explained that, in the WDW context, time is nothing but local correlations between some continuous observable of a semiclassical subsystem and the observables of the quantum subsystem which we are interested in. The resulting time dependence of this quantum subsystem is identical to that prescribed by the time-dependent Schr\"odinger equation.
Similar ideas are discussed in \cite{Lapchinsky:1979fd,Rovelli:1990jm, Rovelli:2009ee, Rovelli:2014ssa,Maniccia:2022iqa}.

We claim that this perspective straightforwardly applies to our quantum mechanical view of the landscape, for example to the toy model landscape with two vacua and a Hamiltonian (cf.~\eqref{alpha_Hilbert_space_structure})
\begin{align}
    {\cal H}_{tot} = {\cal H}_t \oplus {\cal H}_f 
= \Big( \mathcal{H}_{bulk}^t \otimes \mathcal{H}_{horizon}^t \Big) \,\oplus\,
\Big( \mathcal{H}_{bulk}^f \otimes \mathcal{H}_{horizon}^f \Big)\,.\label{H_tot}
\end{align}
The condition that the wave function $\Psi$ solves $H\Psi=0$ ensures that an observer, who is necessarily restricted to measuring in either $\mathcal{H}_{bulk}^t$ or $\mathcal{H}_{bulk}^f$, sees standard quantum mechanical time evolution. There is in particular no need to somehow `save' the scale factor (time) variable $a$ which usually appears in the mini-superspace approach, as we tried to do in Sect.~\ref{tbh}. The reader to whom this is obvious may skip the rest of Sect.~\ref{sect_Banks_approach}.

\subsection{Emergent time in its simplest form}\label{sect_emergent_time}
For a conventional one-particle Hamiltonian of the form
\begin{align}
H_0=\frac{p^2}{2M}+ V(x)\, ,
\end{align}
the familiar WKB solution to the Schr\"odinger equation $H_0\psi_E=E\psi_E$ reads
\begin{align}
\psi_E= [2M(E-V(x))]^{-1/4}e^{\pm i\int_x \sqrt{2M(E-V(x'))}dx'}\, .\label{WKB_approximation}
\end{align}
The WKB method is applicable as long as the potential varies slowly:
\begin{align}
\left|\frac{V'(x)}{(E-V(x))\sqrt{2M(E-V(x))}}\right|\ll 1\, .
\end{align}
Banks \cite{Banks:1984cw} considers a system containing one heavy particle of mass $M$ and a quantum subsystem described by a Hamiltonian $H_l$ ($l$ stands for light degrees of freedom). One then has to solve the WDW equation
\begin{align}
H\Psi = 0 \qquad \mbox{with}\qquad H\,\,=\,\,H_0+H_l \,\,= \,\,\frac{p^2}{2M}+ V(x)+H_l\, .
\end{align}
Here the Hamiltonian $H_l$ also contains a coupling between the heavy particle and the quantum system. Moreover, $M$ is assumed to be large enough to apply WKB and to disregard any backreaction of the quantum system on the heavy particle.

One then makes the ansatz
\begin{align}
\Psi = \sum_E  \alpha_E\psi_E(x)\chi_E(x,y),\quad \alpha_E\in\mathbb{C}\, ,\label{one_particle_ansatz}
\end{align}
where $y$ collectively denotes the positions of the light particles and $\psi_E$ is the WKB solution \eqref{WKB_approximation}. The WDW equation becomes
\begin{align}
\begin{split}
0&=H\Psi = \sum_E \alpha_E \left[E\psi_E\chi_E -\frac{1}{2M}\left(2(\partial_x \psi_E)(\partial_x \chi_E)+\psi_E\partial_x^2\chi_E\right)+\psi_EH_l\chi_E\right]\\
&=\sum_E \alpha_E \left[E\psi_E\chi_E-\frac{1}{2M}\psi_E\left(\pm 2i\sqrt{2M(E-V(x))}\partial_x\chi_E+\partial_x^2\chi_E\right)+\psi_EH_l\chi_E\right]\, .\label{one_particle_calc}
\end{split}
\end{align}
Banks now argues that the term proportional to $\partial_x^2\chi_E$ can be neglected since it is suppressed by a factor of $1/M$. This does not apply to the term involving $\partial_x\chi_E$ since the factor $\sqrt{2M(E-V)}$ may be large. Defining the parameter $t_E$ through a change of variables,
\begin{align}
\frac{dt_E(x)}{dx}=\pm\frac{M}{\sqrt{2M(E-V(x))}}\, ,\label{one_particle_t_def}
\end{align}
the WDW equation takes the form
\begin{align}
0=\sum_E \alpha_E \psi_E\left(-i\partial_{t_E} +E +H_l\right)\chi_E\, .
\label{bseq}
\end{align}
One can now assume that the wave function of the universe is projected on the subspace where the energy of the heavy particle lies in a small interval $(E,E+\delta)$. In practice, this means that \eqref{bseq} is projected to a single, fixed $E$. Intuitively, one may think of the heavy particle being described by a Gaussian wave packet that moves slowly along the real axis. Its position plays the role of time. Thus, solving WDW implies solving the time-dependent Schr\"odinger equation
\begin{align}
i\partial_t\chi_E = (E+H_l)\chi_E\, \label{one_particle_Schroedinger}
\end{align}
for the light degrees of freedom. Here we have replaced $t_E$ by $t$ since the variable $E$ is fixed.
The wavefunction $\chi(t(x),y)$ describes the quantum system of interest relative to the position $x$ of the heavy particle. The additive constant $E$ in \eqref{one_particle_Schroedinger} only contributes a global phase. Time becomes an emergent concept which does not exist outside the validity range of WKB \cite{Banks:1984cw}.

Clearly, the above is just a toy model. One extension would be to a more complicated semiclassical clock, which is certainly doable but non-essential for us. What we consider more important is to gain confidence that the presence of the whole `rest of the world,' in particular the huge horizon Hilbert space, does not invalidate the logic just presented.

\subsection{Including unobserved degrees of freedom}
As just explained, we want to repeat the argument of the previous subsection in the presence of an extremely large additional subsystem which, we assume, is very weakly coupled to our clock and our quantum laboratory. Thus, $\Psi$ is a solution to $H\Psi=0 $ with
\begin{align}
H=H_0+H_l+H_u\, ,
\end{align}
where $H_0$ and $H_l$ are defined as before.
Here $H_u$ (with $u$ for unobserved) governs all the additional degrees of of freedom e.g.~the universe far away from the laboratory and also the much larger space ${\cal H}_{horizon}$. 
In the limit where time scales are not excessively long and where the horizon is only weakly coupled to the bulk system, the total Hilbert space completely factorizes in `unobserved' degrees of freedom and the laboratory system. The total Hamiltonian enjoys a corresponding direct-sum structure. In spite of this factorization, the Hamiltonian $H_u$ contributes non-trivially to the WDW equation viewed as a `zero-energy condition'. The reason is that the `unobserved' degrees of freedom remain in a superposition of states with different energies. Thus, it may not be entirely obvious whether the previously given argument for how standard time evolution follows from the WDW constraint still goes through. In the following, we want to demonstrate that it does. The reader who finds this sufficiently clear at the intuitive level
may jump to Sect. \ref{sect_terminal_vacua}.

Introducing eigenfunctions $\phi_\epsilon(z)$ of $H_u$, i.e. $H_u\phi_\epsilon=\epsilon \phi_\epsilon$ we make the ansatz
\begin{align}
\Psi (x,y,z) = \sum_{E,\epsilon}\alpha_{E,\epsilon}\, \psi_E(x)\chi_{E,\epsilon}(x,y)\phi_\epsilon(z)\, .
\end{align}
Using the result of the previous subsection, the WDW equation becomes
\begin{align}
\begin{split}
0=H\Psi &= \sum_{E,\epsilon}\alpha_{E,\epsilon}\left[\phi_\epsilon H_0(\psi_E\chi_{E,\epsilon})+\psi_E\phi_\epsilon H_l\chi_{E,\epsilon}+\epsilon\psi_E\chi_{E,\epsilon}\phi_\epsilon\right]\\
&=\sum_{E,\epsilon}\alpha_{E,\epsilon}\, \psi_E\, \phi_\epsilon\left(-i\partial_{t_E} +E+\epsilon+H_l\right)\chi_{E,\epsilon}\, ,\label{WDW_multi_unob}
\end{split}
\end{align}
Assuming, as before, that the energy $E$ of the clock system is known with sufficient precision, we find that the wave function of the universe is
\begin{align}
\Psi = \sum_\epsilon\alpha_\epsilon\psi_E\chi_{E,\epsilon}\phi_\epsilon\, .
\label{3fa}
\end{align}
Here the wave functions $\phi_\epsilon$ determine the state of the unobserved part of the universe while, crucially, $\chi_{E,\epsilon}$ are solutions of the time-dependent Schr\"odinger equation
\begin{align}
i\partial_t\chi_{E,\epsilon}=\left(E+\epsilon+H_l\right)\chi_{E,\epsilon}\,.
\label{gen_S_eq}
\end{align}

At this point, one might be concerned that our wave function of the universe involves a superposition of $\chi_{E,\epsilon}\phi_\epsilon$ with different $\epsilon$, such that the `unobserved' part of the universe may after all affect our measurements. However, this does not happen for the following reason: Note first that solutions of \eqref{gen_S_eq} are given by
\begin{align}
\chi_{E,\epsilon}(t,y)=e^{-i(E+\epsilon)t}\chi (t,y)\, ,\label{psep}
\end{align}
where $\chi (t,y)$ satisfies the Schr\"odinger equation for the Hamiltonian $H_l$:
\begin{align}
i\partial_t \chi (t,y)=H_l\chi (t,y)\, .
\end{align}
It is now easy to see that the phase in \eqref{psep} is irrelevant for the type of observables we care about. Indeed, such a typical observable is
\begin{align}
O=O_0\otimes O_l\otimes\mathbb{1}_u\,,
\label{physical_operator}
\end{align}
where $\mathbb{1}_u$ is the identity acting on ${\cal H}_u$. In addition, $O_l$ is a generic observable of our quantum system with the light degrees of freedom. Finally $O_0$ is, without loss of generality, a function of $x$, which is equivalent to a function $f(t)$ of our time variable $t$.
Recalling \eqref{3fa} and \eqref{psep}, it is now easy to see that
\begin{align}
\begin{split}
\bra{\Psi}O\ket{\Psi}&=\sum_{\epsilon,\epsilon'}\alpha_{\epsilon'}^*\alpha_\epsilon e^{+i\epsilon't}e^{-i\epsilon t}\,f(t)\,\bra{\chi (t)}O_l\ket{\chi (t)}\,
\bra{\phi_{\epsilon'}}\mathbb{1}_u\ket{\phi_\epsilon}
\\
&=\left(\sum_\epsilon |\alpha_\epsilon |^2\right) \,f(t)\,\bra{\chi (t)}O_l\ket{\chi (t)}\,.
\end{split}
\end{align}
We see that, as promised, the unobserved part of our Hilbert space is irrelevant in that it affects only the constant prefactor. Apart from this prefactor, we obtain the desired time-sensitive measurement of $O_l$ on the basis of the solution $\chi(t)$ of our time-dependent Schr\"odinger equation.

\subsection{The complete picture of a universe without time}
To summarise, we have recalled and slightly generalized Bank's argument about the emergence of the time-dependent Schr\"odinger equation.
For us, this is a strong motivation to believe that the emergence of time is consistent with (but does not rely on) the presence of a huge, unobserved Hilbert space, especially that `behind the horizon.' Thus, we do not create any conceptual problems when working exclusively with the WDW equation $H\Psi=0$, without any scale-factor-related or other time variable.

At this point, one might worry that for a generic Hamiltonian acting on a finite-dimensional Hilbert space, the WDW equation will have no solution.
This issue will be resolved in Sect.~\ref{sect_terminal_vacua} when more realistic landscapes including terminal vacua are considered.
The landscape with two dS vacua discussed until now should only be seen as a toy-model.
The solution to the WdW equation is a fundamentally time independent state $\Psi$ that encodes all information about the universe. The probabilities of results of certain observations, labelled by $i$, are simply proportional to $\Vert P_i\Psi\Vert^2$. Here $P_i$ are appropriate projection operators projecting $\Psi$ on subspaces where the result $i$ is guaranteed and $\Vert\cdot\Vert$ is the Hilbert space norm of the finite-dimensional space \eqref{H_tot}. No averaging over time is needed.
As an example, the probability that a random observer sees vacuum $t$ or $f$ is now \textit{by definition} given by $\Vert \Psi\Vert_{t,f}^2$. We may finally compare this result with the discussion of Sect.~\ref{ede} and, in particular, Eq.~\eqref{ratio_expr}. From this, we conclude that the WDW approach together with the assumption of quantum ergodicity (cf.~\eqref{Shnirelman_ratio}, as motivated by Shnirelman's theorem) agrees with semiclassical expectations.

\section{Terminal vacua}\label{sect_terminal_vacua}
As we have just seen, the model based on \eqref{Hamilton} and the WDW equation $H\Psi=0$ leads to predictions consistent with semiclassical physics.
This straightforwardly generalizes from our toy-model landscape with two vacua, labelled by $t$ and $f$ for true and false, to a landscape with multiple dS vacua. Labelling the vacua by an index $i$, we now assume that the full Hilbert space is of the form
\begin{align}
    \mathcal{H}=\bigoplus_{i}\mathcal{H}_i\,,
\end{align}
with $N_i\equiv \text{dim}(\mathcal{H}_i)=\exp(S_i)$, and with the Hamiltonian being approximately block diagonal, with the off-diagonal blocks inducing tunneling. This is the natural generalization of \eqref{Hamilton}. The Shnirelman theorem and random matrix theory now suggest that 
\begin{align}
    \Vert \Psi\Vert_i^2=\frac{N_i}{\sum_{j} N_j}\, ,\label{Shnirelman_extended}
\end{align}
where $\Vert\cdot\Vert_i$ is the norm of $\mathcal{H}$ restricted to ${\cal H}_i$. This again agrees with probabilities derived from rate equations based on semiclassical tunneling  \cite{Vanchurin:2006qp}.

After these preliminaries, we now turn to the main and less straightforward subject of the present section: The inclusion of AdS vacua. We will assume that AdS vacua are terminal, in other words, we do not consider the possibility that the dynamics of a big crunch induces tunneling processes back to a dS vacuum. One may motivate this by noting the similarity between crunching AdS space and the interior of a black hole near the singularity. 
For black holes, it is widely believed that they evaporate in a unitary process and hence without transitions to other universes. We then assume the same for AdS crunches.

However, we would not claim that the analogy above is fully convincing.
For us, it is merely a working assumption. We do not want to exclude, as a matter of principle, the interesting possibility that the large energy densities of a big crunch induce tunneling to dS, as explored e.g.\ in \cite{Piao:2004me,Johnson:2011aa,garriga2013watchers,Garriga:2013cix,Gupt:2013poa,Liu:2014uda}.
It will become clear below that such processes can be straightforwardly implemented in the model we are about to construct.

\subsection{Terminal vacua as infinite-\texorpdfstring{$N$}{N} limits}
One possibility for including a terminal vacuum (AdS or Minkowski) in the WDW framework is to add an infinite-dimensional subspace ${\cal H}_{ter}$ and, correspondingly, to extend the Hamiltonian by a block of infinite size. To implement this, we first extend ${\cal H}$ by another block of size $N_{ter}\times N_{ter}$ and the corresponding off-diagonal blocks responsible for tunneling. We first assume $N_{ter}$ to be very large but finite: $N_{ter}\gg N_i$ for all $i$. Inspired by ergodic arguments, one again expects
\begin{align}
    \Vert \Psi\Vert_{i,ter}^2=\frac{N_{i,ter}}{N_{ter}+\sum_j N_j}\, .
\end{align}
This means that ratios of the form $\Vert \Psi\Vert_i^2/\Vert \Psi\Vert_j^2$ evaluate to $N_i/N_j$, as before. If we disregard AdS and Minkowski vacua as anthropically unsuitable,\footnote{
In spite of the big crunch arising in AdS and the infinite, cold and empty future of Minkowski space, this is not an obvious assumption. Observers like ourselves could, of course, nevertheless exist in an initial phase after the last tunneling event. However, this opens up the non-trivial question of how to obtain probabilities specifically for such `inflationary' observers. We will comment on this in Sect.~\ref{sect_applications}
} 
the dominance of the terminal block, $\Vert \Psi\Vert_{ter}^2\gg \Vert \Psi\Vert_i^2$, plays no role. 
Now we can safely take the limit $N_{ter}\to \infty$.
Clearly, the full state of the multiverse will not be normalizable in this limit, but since we are only interested in ratios between the probabilities for different dS vacua, this is not an issue.

Unfortunately, this suggested treatment of terminal vacua gives rise to a significant concern: We are allowing for a non-zero transition amplitude from dS to the terminal vacuum (or many terminal vacua). At the same time, we are looking either for a stationary solution (in the naive approach) or for a fundamentally time-independent WDW wave function. As we explained, the latter may then be reinterpreted by introducing an emerging semiclassical time variable. Either way, since we are not leaving the framework of conventional quantum mechanics, our stationary solution must be based on a non-zero probability both for tunneling events from dS to terminal vacua {\it and} for some form of transition from terminal vacua back to dS. For the latter, we have no semiclassical understanding (see, however, \cite{nomura2012static, DeAlwis:2019rxg}). One could, of course, simply accept this and postulate that such AdS-to-dS transitions are nevertheless possible, making our approach self-consistent. For example, one may hope that a semiclassical understanding will be achieved in the future or one may assume that such processes exist but do not allow for a semiclassical description. However, we prefer to take a more conservative point of view and look for a different and maybe more natural resolution of the difficulty just described.

\subsection{The creation of dS vacua and probability currents}\label{cds}
As advertised, we now attempt to develop our approach such that terminal vacua are allowed and semiclassical intuition is nevertheless obeyed. 
To find a time-independent wave function of the universe without tunneling processes from terminal to dS vacua, one then also needs a non-vanishing amplitude for dS space to be created out of nothing. One may think of a 3-sphere being created at zero radius and then growing. The euclidean version of this process is known as the Hartle-Hawking no-boundary proposal \cite{hartle1983wave}. We will be agnostic concerning the value of the corresponding amplitudes, for which different proposals associated with the names of Hartle-Hawking \cite{hartle1983wave}, Linde \cite{Linde:1983mx} and Vilenkin \cite{Vilenkin:1984wp,Vilenkin:1986cy} exist. In either case, a source term appears in the WDW equation \cite{Halliwell:1988wc, Teitelboim:1981ua}, making the wave function of the universe a Green's function to the Hamiltonian.

To illustrate the above, let us start by discussing a toy model entirely unrelated to dS space or cosmology: Consider the time independent Schr\"odinger equation for a particle on the real line $\mathbb{R}\ni x$ and add an inhomogeneous term, governed by a constant $c$ and localized at the origin:
\be
(\partial_x^2+E)\psi(x)=2ic\delta(x)\,.
\ee
One immediately concludes that $\partial_x\psi$ jumps by $2ic$ at $x=0$ and hence
\begin{align}
	\begin{split}
			\psi(x)&=(c/k)\exp(ikx)\quad\mbox{for}\quad x > 0\, ,\\
			\psi(x)&=(c/k)\exp(- ikx)\quad\mbox{for}\quad x< 0\quad,\qquad\mbox{with}\quad k=\sqrt{E}\,\,,
		\end{split}
\end{align}
represents a solution.
The familiar quantum mechanical probability current $j(x)$ for this wavefunction is given by
\begin{align}
		\begin{split}
			j(x)&= |c|^2/k\quad\mbox{for}\quad x > 0\, ,\\
			j(x)&= -|c|^2/k\quad\mbox{for}\quad x < 0\, .
		\end{split}
\end{align}
Because $\nabla\cdot j(x)=2|c|^2/k\, \delta(x)$, the solution corresponds to a source at $x=0$.
Similarly, we find a second solution with $k=-\sqrt{E}$ where two waves coming from infinity end in a sink at $x=0$.
Both solutions exist and it is a question of boundary conditions which solutions to use.
We see that the source triggers a probability current which can, due to the unboundedness of the real line, flow to infinity and thus realize a stationary solution without the need for a sink or any current coming back from infinity.
For a particle localized in an infinitely steep square well, this would be impossible as the energy eigenfunctions are standing waves, leading to a vanishing probability current.
	
We now want to transfer the above intuition to the multiverse, where the Hartle-Hawking or Linde/Vilenkin mechanism for the creation of dS spaces out of nothing sources the probability flux. 
As a first step, we need to define a probability current for systems that are more general than a particle in $\mathbb{R}^n$. Such a generalization, using only the existence of a countable Hilbert space basis $\{\ket{n}\}$, is known \cite{de1992probability,schumacher2016probability,Roden_Probability_current}: In this setting, the current is a matrix 
	\begin{align}
		J_{nm}=-i\left(H_{nm}\rho_{mn}-\rho_{nm}H_{mn}\right)\, ,\label{Current_definition}
	\end{align}
where $H_{nm}=\bra{n}H\ket{m}$ and $\rho_{nm}=\bra{n}\rho\ket{m}$, with $\rho$ being the density matrix.
Since we are working with a pure state, $\rho = \ket{\Psi}\bra{\Psi}$. The following formulas straightforwardly generalize to more general density matrices.
The matrix $J$ is real and antisymmetric and its elements $J_{nm}$ measure the probability flowing from $\ket{m}$ to $\ket{n}$.In an ordinary quantum mechanical system with standard time evolution, which for the moment serves as a tool to build intuition, $J$ obeys the relation
	\begin{align}
		-\partial_t |\braket{n}{\Psi}|^2 =-\sum_m J_{nm}=\sum_m J_{mn}\, .\label{discrete_continuity_equation}
	\end{align}
The r.h.~side of this equation quantifies the rate with which probability flows out of the state $\ket{n}$, similar to the divergence $\nabla\cdot j(x)$ in the continuum case. Equation \eqref{discrete_continuity_equation} represents the discrete analogue of the continuity equation.
In the continuum limit, the definition \eqref{Current_definition} coincides with the familiar probability current of point-particle quantum mechanics \cite{schumacher2016probability}. We reiterate that we have spelled out the effect of $J_{mn}$ on time evolution in \eqref{discrete_continuity_equation} exclusively for better intuition and the rest of this section is devoted to the stationary case.
	
For stationary solutions to the Schr\"odinger equation, the l.h.~side of \eqref{discrete_continuity_equation} is zero and hence the r.h.~side also vanishes. This changes if $\Psi$ is a solution to the WDW equation in the presence of a source. However, as long as the Hilbert space remains finite-dimensional, sources always come together with sinks. To show this note that, by antisymmetry, $\sum_n(\sum_m J_{mn})=0$. This implies that the divergences of the probability currents at $n$, summed over all $n$, give zero. Thus, sources without sinks are impossible.
The process of creating dS vacua out of nothing has no inverse, i.e. dS vacua can not decay into `nothing' and hence they do not act as sinks.
We immediately conclude that \textbf{a landscape containing only dS vacua is inconsistent}. The problem disappears if one includes terminal vacua with their infinite-dimensional Hilbert spaces since now the probability current can run to infinity, cf. our toy-model above. Terminal vacua become a necessary part of the landscape. The full picture is that there is a probability flow from dS to terminal vacua, even though the state of the universe is a solution to the WDW equation and thus static.

It is well known that only dS vacua can be created out of nothing. As a result, the source is a vector from the dS-part of the total Hilbert space. The WDW equation including the source takes the simple form
\begin{align}
   H\Psi=\chi\, ,\label{WDW_source_Matrix}
\end{align}
where $\chi$ is a vector from the dS part of ${\cal H}$ (i.e.~its projection on terminal vacua vanishes).
If $\chi\perp \mbox{ker}(H)$, which we will from now on assume, then solutions to \eqref{WDW_source_Matrix} exist and take the general form
\begin{align}
	\Psi=\Psi_0+\Psi_s\, .\label{hom_and_inhom_solution}
\end{align}
 Here the first term is a solution to the homogeneous WDW equation, $\Psi_0\in\mbox{ker}(H)$, and the second term denotes a special solution.
We would like to fix $\Psi$ by imposing appropriate boundary conditions. As discussed before, the solution to the WDW equation should be such that the probability current runs out to infinity without coming back. This is equivalent to requiring that the vector $\chi$ acts purely as a source and not as a sink. Hence, our boundary conditions are that there is no probability current coming from infinity in the Hilbert spaces of the terminal vacua.
This is analogous to our toy-model above where we have chosen the solution where the inhomogeneous term acts as a source. One may hope that these conditions are sufficient to fix the state of the universe. It now becomes clear why we do not expect transitions from terminal to dS vacua, even though an off-diagonal matrix elements in the Hamiltonian exists (which is responsible for the familiar down-tunneling process): The Hartle-Hawking or Linde/Vilenkin proposal requires dS vacua to act as sources and not as sinks, leading to the consideration of boundary conditions forbidding a current coming from infinity.

Assuming that $\Psi$ is uniquely determined, the relative probability for a random observer in the multiverse to be in dS vacuum $i$ is then given by
\begin{align}
p_i\equiv \Vert \Psi\Vert_i^2\, .\label{probability_pj}
\end{align}
Since we know neither the precise form of the source vector nor the microscopic Hamiltonian, practical calculations are problematic.
In the pure dS landscape without sources (Sect.~\ref{nvi}), the Shnirelman theorem was a helpful tool to calculate the probability distribution of vacua and, more importantly, to gain confidence in the quantum-mechanical point of view.
In the more complicated setting we are considering at present, where terminal vacua and sources are included, the Shnirelman theorem (at least in the form \eqref{Shnirelman_extended}) is not expected to hold.
However we can make progress using semi-classical results. To explain this, let us collect some facts concerning our probability current:

The strength of a source is measured by the divergence of the probability current, as on the r.h. side of \eqref{discrete_continuity_equation}. Consider now a dS vacuum $i$ with basis $\{\ket{n}\}$ and denote by $\{\ket{\alpha}\}$ a basis of the full Hilbert space. The strength of the source in this dS space then reads
\begin{align}
J_i\equiv \sum_{\alpha, n}J_{\alpha n}\, .\label{Total_divergence}
\end{align}
Consider a second vacuum $x$ (dS or terminal) with basis $\{\ket{r}\}$. The total probability flow from vacuum $i$ to $x$ reads
\begin{align}
	J_{i\to x}\equiv\sum_{r,n}J_{rn}=2\,{\rm Im}\left(\sum_{r,n}\bra{r}H\ket{n}\bra{n}\ket{\Psi}\bra{\Psi}\ket{r}\right)\, .
	\label{jix}
\end{align}
Here, we encounter projection operators $P_i=\sum_n \ket{n}\bra{n}$ and $P_x=\sum_r\ket{r}\bra{r}$ on the subspaces belonging to vacua $i$ and $x$ respectively.
Using the shorthand notation $\ket{\Psi}_{i,x}\equiv P_{i,x}\ket{\Psi}$, we find
\begin{align}
J_{i\to x}=2\,{\rm Im}\left(\,{}_x\!\bra{\Psi}H\ket{\Psi}_i \,\right)\, .\label{Total_Flow}
\end{align}
This expression is independent of the chosen basis and represents the fact that the total flow between the vacua $i$ and $x$ is governed by the off-diagonal matrix elements of the Hamiltonian.
	
Consider now the case when $x$ is a dS vacuum. Then, the probability current is linked to the tunneling rates $\Gamma_{i\to x}$ and $\Gamma_{x\to i}$ between the two vacua via
\begin{align}
    J_{i\to x}=p_i\Gamma_{i\to x}-p_x\Gamma_{x\to i}\, .\label{J_link_Gamma_dS}
\end{align}
Given \eqref{jix}, we could now determine the decay rates $\Gamma$ on the r.h.~side from the fundamental input data $\chi$ and $H$. The latter is an appropriate generalization of our block-diagonal ansatz for the Hamiltonian in \eqref{Hamilton}.
    
If $x$ defines a terminal vacuum, the chosen boundary conditions constrain the probability current to the form
\begin{align}
    J_{i\to x}=p_i\Gamma_{i\to x}\, .\label{J_link_Gamma_AdS}
\end{align}
Again, given $H$ and $\chi$ we could in principle compute the rates $\Gamma$.
    
However, as advertised before, we do not understand the fundamental input data well enough and it is more practical to turn the logic around: We can start with our semiclassical knowledge of (or proposals for) the source strengths $J_i$ and CDL rates $\Gamma$ \cite{coleman1980gravitational}.\footnote{
In 
fact, by identifying the transitions rates $\Gamma$ on the r.h.~side of \eqref{J_link_Gamma_dS} with CDL rates, we are probably making a (non-exponential) error associated with the precise definition of the volume over which our CDL rate is to be integrated and possibly with further technical details.
} 
Then the probabilities $p_i$ can be derived without knowledge of $\Psi$.
    
Let us start with the $J_i$, for which the two key competing proposals are
\begin{align}
    J_i\propto \begin{cases} \exp(S_i) & \text{no-boundary \cite{Hawking:1998bn}}\\ 
        \exp(-S_i) & \text{tunneling \cite{ Linde:1983mx, Vilenkin:1984wp}}
        \end{cases}\, .\label{source_strength}
    \end{align}
Clearly, by the definitions \eqref{Total_divergence} and \eqref{Total_Flow}, they satisfy
    \begin{align}
        J_i=\sum_x J_{i\to x}\, ,
    \end{align}
where the summation is over all vacua of the landscape. Using \eqref{J_link_Gamma_dS} and \eqref{J_link_Gamma_AdS}, one finds the set of equations (one for each dS vacuum $i$)
\begin{align}
    J_i=\sum_{j\,\in\,dS} \left(p_i\Gamma_{i\to j}-p_j\Gamma_{j\to i}\right)+p_i\sum_{y\,\in\, Terminals}\Gamma_{i\to y}\, ,\label{defining_equation}
    \end{align}
where $j$ runs over all dS vacua and $y$ over all terminals. This can be read as matrix equation, $J_i=M_{ij}p_j$, and hence easily solved:
$p_i=(M^{-1})_{ij}J_j$.\footnote{
The 
matrix $M$ also appears in \cite{Garriga:2005av} where it was shown that all eigenvalues are nonzero and hence the matrix is invertible.
} 
It is interesting to note that the normalization of the $J_i$ does not play a role. A rescaling of the $J_i$ can be absorbed in the $p_i$, leaving only relative probabilities of the form $p_i/p_j$ as physical predictions.
This makes sense also intuitively since we do not expect to be able to derive the precise normalization of the production rates $J_i$.

We note that the results \eqref{WDW_source_Matrix} and \eqref{defining_equation} are equally meaningful in the absence of a traditional de Sitter landscape. If the underlying quantum gravity theory is such that only solutions with a decaying dark energy are available \cite{Danielsson:2018ztv, Ooguri:2018wrx, Bedroya:2019snp, Bedroya:2019tba, Rudelius:2019cfh, Wang:2019eym, Blanco-Pillado:2019tdf, Brahma:2020cpy}, then on the r.h.~side of \eqref{defining_equation} the index $j$ runs only over such relatively short-lived vacua with temporal acceleration. As a result, the equation will be dominated by the source terms and the decays to terminals.
Anthropic predictions as discussed in Sect.~\ref{sect_applications} can nevertheless be made with the same methods.
It is also clear how to modify the model when transitions from crunching AdS regions to dS spaces are allowed. One simply includes the corresponding probability current in \eqref{defining_equation}. 
However, since such events are UV sensitive, the form of the probability current is unknown.

Similar rate equations have appeared before in the discussion of the scale-factor \cite{Linde:1993nz,DeSimone:2008bq,DeSimone:2008if,Garriga:2005av} and watcher \cite{garriga2013watchers} measure, but in conceptually different frameworks. It is particularly interesting that, if one identifies our $p_i$ with the relative time an eternal observer spends in vacuum $i$, then an (almost) equivalent rate equation including sources can be found in \cite{garriga2013watchers}. There, however, it is assumed that the observer somehow `survives' the crunch in the terminal AdS vacuum and returns to one of the newly-born dS vacua, with a relative probability given by the no-boundary or tunneling proposal. While it is intuitive that this type of dynamics leads to $p_i$s which agree with our result, we feel that conceptually our treatment is rather different: We require no observer and no assumption about any return from a terminal vacuum. This ensures in particular that SUSY-Minkowski vacua, in which there are no further decays and no big crunch (presumably implying no possibility to return), do not require any special treatment in our approach. By contrast, the main proposal of~\cite{garriga2013watchers} is to restrict attention to the measure-zero subset of observer trajectories which never enter a Minkowski vacuum.

We note, however, that one of the different possibilities for defining a measure discussed in~\cite{garriga2013watchers} is, in our present understanding, equivalent to our result at the technical level: In this approach, it is assumed that the observer trajectories truly end in either eternal Minkowski or crunching AdS, such that `returns' of the observers are replaced by an a priori choice of a certain ensemble of observer trajectories. If this choice agrees with what we implement using the fundamental source of the WDW equation, then predictions should agree. 

Finally, a similar rate equation also appears in \cite{Garriga:2005av}, where the $p_i$ represent the time-averaged fraction of comoving volume occupied by vacuum $i$ and the $J_i$ represent the initial distribution of vacua.
The analysis in \cite{Garriga:2005av} is based on the conventional picture of the multiverse having infinite volume and containing infinitely many bubbles of each kind.
The initial condition dependence of the resulting equation then appears to be counter-intuitive to the authors.
The local WDW measure on the other hand naturally does not consider an infinite-volume spacetime geometry and the inhomogeneous version of the WDW equation \eqref{WDW_source_Matrix} leads to a constant probability flux responsible for the permanent creation of dS vacua out of nothing.
Hence, it is by no means unexpected that the distribution of vacua depends on the creation rates of dS spaces.

Ultimately, it may not seem surprising that we find a result already predicted by one of the many measures based on geometric cutoffs.
But in our approach the resulting measure is not just one of many options -- we claim it to be the unique outcome of combining quantum mechanics, semiclassical gravity, and the cosmological central dogma.

\section{Applications and comments on related approaches}
\label{sect_applications}
To illustrate our measure proposal, we will now analyze the probability distribution of vacua in a toy model landscape. We will then comment on the more complicated question where post-inflationary observers like ourselves are expected to live.

\subsection{A toy-model landscape}\label{sect_toy_landscape}
Consider two dS vacua labeled `$1$' and `$2$' together with a terminal vacuum labeled `$T$.' 
Without loss of generality, let $S_1<S_2$.
We are interested in finding $p_1,p_2$ by solving equation \eqref{defining_equation}.
In our case of interest, this equation reads
\begin{align}
	\begin{pmatrix}
		J_1\\
		J_2
		\end{pmatrix}&=\begin{pmatrix}
		\Gamma_{1\to 2}+\Gamma_{1\to T} & -\Gamma_{2\to 1}\\
		-\Gamma_{1\to 2} & \Gamma_{2\to 1}+\Gamma_{2\to T}
	\end{pmatrix}\begin{pmatrix}
	p_1\\
	p_2
    \end{pmatrix}\,,
    \label{req}
\end{align}
which is solved by
\begin{align}
	\begin{pmatrix}
			p_1\\
			p_2
		\end{pmatrix}=\frac{1}{\det(M)}\begin{pmatrix}
		\Gamma_{2\to 1}+\Gamma_{2\to T} & \Gamma_{2\to 1}\\
		\Gamma_{1\to 2} & \Gamma_{1\to 2}+\Gamma_{1\to T}
	\end{pmatrix}\begin{pmatrix}
	J_1\\
	J_2
    \end{pmatrix}\,.
    \label{reqi}
\end{align}
Here $M$ denotes the transition rate matrix from \eqref{req}. Let us make the further simplifying assumption that tunneling to the terminal vacuum is always much faster than tunneling to the other dS vacuum. This is in line with stringy expectations due to the well-known difficulties in constructing dS vacua. The latter are hence expected to be rare in a string landscape consisting of mostly AdS vacua. Transitions between dS vacua are then likely to involve changing the flux by many quanta or even the topology, suppressing the decay rates. We hence approximate the diagonal components of the matrices in \eqref{req} and \eqref{reqi} by just $\Gamma_{1\to T}$ and $\Gamma_{2\to T}$.

\paragraph{Linde/Vilenkin:} The tunneling proposal \cite{Linde:1983mx, Vilenkin:1984wp, Vilenkin:1986cy} for the creation of dS vacua suggests $J_i\propto \exp(-S_i)$ and hence
\begin{align}
    \frac{p_1}{p_2}\simeq \frac{\Gamma_{2\to T}e^{-S_1}+\Gamma_{2\to 1}e^{-S_2}}{\Gamma_{1\to 2}e^{-S_1}+\Gamma_{1\to T}e^{-S_2}}\simeq \frac{\Gamma_{2\to T}e^{-S_1}}{\Gamma_{1\to 2}e^{-S_1}+\Gamma_{1\to T}e^{-S_2}}\,.
\end{align}
There are now several possibilities: Depending on which of the two large ratios $e^{S_2}/e^{S_1}$ and $\Gamma_{1\to T}/\Gamma_{1\to 2}$ wins, either the first or the second term in the denominator dominates. If the first term dominates, $p_1 \gg p_2$. If the second term dominates, one has to distinguish two further cases. Now the competition is between the large ratio $e^{S_2}/e^{S_1}$ and the ratio $\Gamma_{1\to T}/\Gamma_{2\to T}$, which could in principle also be exponentially large. If it is and if it wins over $e^{S_2}/e^{S_1}$, then $p_2\gg p_1$. Otherwise we have $p_1\gg p_2$, as before.

\paragraph{Hartle-Hawking:} 
By contrast, the no-boundary proposal \cite{hartle1983wave} states that $J_i\propto \exp(S_i)$, leading to 
\begin{align}
    \frac{p_1}{p_2}\simeq \frac{\Gamma_{2\to T}e^{S_1}+\Gamma_{1\to 2}e^{S_1-S_2}e^{S_2}}{\Gamma_{1\to 2}e^{S_1}+\Gamma_{1\to T}e^{S_2}}\simeq \frac{\Gamma_{2\to T}e^{S_1}}{\Gamma_{1\to T}e^{S_2}}\,.
\end{align}
The ratio of probabilities is determined by the competition between the large ratio $e^{S_2}/e^{S_1}$ and the ratio $\Gamma_{2\to T}/\Gamma_{1\to T}$. Since vacuum 1 has the higher energy scale, one would naively expect that it decays more easily, in which case both the entropy factors and the decay rates favor the second vacuum: $p_2\gg p_1$. However, one can not completely rule out that the tensions of the relevant domain walls lead to $\Gamma_{2\to T}/\Gamma_{1\to T}\gg e^{S_2}/e^{S_1}$, implying the opposite result.

\paragraph{Preliminary conclusions:}
In the absence of large hierarchies between the decay rates to the terminal vacuum, the dS space which is more likely to be created out of nothing has a significantly higher probability of being realized in the multiverse. Thus, as naively expected, high-scale vacua or low-scale vacua win depending on whether the Linde/Vilenkin or Hartle-Hawking proposal is correct. However, this expectation can be overturned by large ratios of the decay rates, especially those from the relevant dS to the terminal vacuum. A much more careful analysis, including the realistic case of a large number of dS and terminal vacua and using what is known about decay rates in the string landscape, is clearly required to make progress.

\subsection{The most likely vacuum for an anthropic observer}
\label{sect_anthropic_prediction}
The ultimate goal in analysing the measure problem is, of course, to make phenomenological predictions in an eternally inflating universe. In addition to the famous cosmological constant issue \cite{Weinberg:1987dv} (cf.~\cite{Bousso:2007gp} for a more recent discussion and references), the targets of such analyses include the curvature parameter $\Omega_k$ \cite{Freivogel:2005vv, DeSimone:2009dq, Guth:2012ww}, the tensor-to-scalar ratio \cite{Westphal:2012up, Pedro:2013nda}, and the SUSY breaking scale \cite{Douglas:2012bu}.\footnote{See \cite{Baer:2022naw} for a recent review emphasizing the alternative perspective of `stringy naturalness.' The latter is based on counting which features occur more frequently in the string landscape.}

In this subsection, we do not attempt a serious phenomenological study but merely want to discuss the first steps required to go from the distribution of vacua per se to the prediction in which vacuum an anthropic observer is most likely to live.
Here by anthropic observers we mean some intelligent beings like us, which developed on the basis of structure formation in the universe (as opposed to 
Boltzmann Brains or `abstract observers' invented to count vacua). 
It is well-understood that anthropic observers can only develop under special conditions within the set of universes and initial conditions consistent with fundamental theory,  and that requiring these conditions leads to strong and robust predictions on observable quantities. The most basic constraint is that the cosmological constant \cite{Weinberg:1987dv} must be small enough to allow for the formation of gravitationally bound structures. Structure formation can only take place following slow-roll inflation when dark matter abundance and strength of primordial fluctuations are in a certain range \cite{Linde:1987bx,Linde:1991km,Hellerman:2005yi}. Eventually, biological evolution must be taken into account. As a result, the conditions on the cosmological parameters above become even more stringent and further cosmological and particle-physics constraints arise \cite{Tegmark:2005dy}.

To set the stage, recall that when using the scale factor measure the landscape is populated mainly through the so-called dominant (or master) vacuum \cite{Garriga:2005av, Schwartz-Perlov:2006swo}. This is the dS vacuum with the smallest decay rate. It generally has a small cosmological constant, though not necessarily the smallest one (since decay rates also depend on other parameters).
Thus, anthropic observes are most likely to live in vacua to which the tunneling rate from the dominant vacuum is high. More precisely, the vacua in question have to be adjacent to an inflationary plateau and it is the tunneling rate from the dominant vacuum to this plateau that has to be high.

In our case, where the multiverse is characterized by a quantum state $\Psi$ which crucially depends on the possibility of creating dS vacua out of nothing, the situation is different. It is in general not a good approximation to disregard every but the dominant vacuum at late times. The reason is that all vacua are populated by creation out of nothing or by subsequent down-tunneling events. This can easily be more important than the up-tunneling from the dominant vacuum, which is double-exponentially suppressed (by the bounce-action as well as by the large entropy of the dominant vacuum). Thus, the entire landscape will constantly be populated and up-tunneling events may not be particularly important.

On the basis of a stationary state $\Psi$ of the universe, (relative) probabilities for measurements can be obtained by projecting $\Psi$ on the subspace of interest and then taking its norm squared. Without observers and assuming that the measurement simply determines the type of vacuum $i$, one would find the probabilities $p_i$ of Sects.~\ref{cds} and \ref{sect_toy_landscape}. But the projector we are really interested in is on the vacuum $i$ and, on top of that, on the observer in that vacuum. This may be taken into account by an additional weighting factor $w_i$. The probability $O_i$ that an anthropic observer lives in dS vacuum $i$ is then given by
\begin{align}
	O_i\propto w_ip_i\,,
\end{align}
where the $p_i$ follow from the local WDW measure by solving equation \eqref{defining_equation}. We do not want to enter the non-trivial discussion of how to implement the projection in cases where typical spatial slices of the relevant dS static patch contain multiple anthropic observers. Naively one would expect that this leads to large $w_i$ and hence rewards, for example, vacua with a large horizon size (i.e.~small cosmological constant). However, we do not expect hierarchies between the $w_i$ to be exponential.\footnote{
If 
post-inflationary observers can live in $\mathcal{N}=2$ Minkowski vacua, this assumption is problematic. Indeed, in this case one of the possible projections on $\Psi$ would be
an infinite reheating surface in the relevant vacuum. 
} 
Thus, since we have disregarded exponential effects throughout, we may for the sake for concreteness take $w_i\in\{0,1\}$, classifying all vacua of the landscape as anthropically suitable or not suitable.

However, as we have argued before, anthropic observers live after a period of inflation, which only occurs shortly after the creation of a pocket universe belonging to vacuum $i$. Hence, a more appropriate estimate for $O_i$ should be based on the production rate of inflationary universes of vacuum $i$ and not on the abundance of vacuum $i$ in the multiverse. 
A corresponding ansatz is
\begin{align}
	O_i \propto w_i \left( f_i J_i + \sum_{j\neq i}p_jf_{ji}\Gamma_{j\to i}\right)\,,\label{O_j}
\end{align}
where $J_i$ denotes the creation rate of vacuum $i$ from nothing and $\Gamma_{j\to i}$ the transition rate from vacuum $i$ to vacuum $j$. In addition, $f_i$ and $f_{ji}$ are the fractions of creation and tunneling events that end specifically in an inflationary state of vacuum $i$ rather than anywhere within the basin of attraction of the minimum of the scalar potential describing vacuum $i$. Any attempts to derive $f_i$, $f_{ij}$ in the known part of the string landscape are left to future work. We note that, since the $J_i$ and $p_i$ are defined only up to arbitrary overall rescalings, eq.~\eqref{O_j} implies that only ratios of the form $O_i/O_j$ are physically meaningful.

Most naively, one may be tempted to assume that the index $i$ in \eqref{O_j} runs over dS vacua only. However, it takes little thought to see that also Minkowski vacua or AdS spaces with near-zero cosmological constant should be included. Indeed, such vacua may have adjacent inflationary plateaus which can be populated by tunneling or creation out of nothing. Slow roll and reheating can then end in the terminal vacuum and structure formation can occur long before the big crunch (in the AdS case).
To summarize, eq.~\eqref{O_j} holds also when vacuum $i$ is terminal. It is then understood that $f_iJ_i$ stands for the creation rate of a quasi-dS vacuum (i.e.~an inflationary state) leading to reheating in vacuum $i$.

Before closing this section, we would like to discuss an alternative point of view: Slow-roll inflationary plateaus could by definition be promoted to additional landscape vacua. Let us label such vacua by the index `$i,inf$' if the slow-roll ends in the metastable dS or terminal vacuum $i$. The probability for an anthropic observer to be in vacuum $i$ is then proportional to $w_i p_{i,inf}$, where $p_{i,inf}$ is the probability that the universe is in the `vacuum' defined by the inflationary plateau near minimum $i$. The $p_{i,inf}$ still follow from \eqref{defining_equation}, where appropriate creation rates of and tunneling rates to vacua of type $p_{i,inf}$ have to be added. Moreover, decay rates of such quasi-dS vacua also have to be introduced. These are dominated by the slow-roll/reheating transition to vacuum $i$ rather than by tunneling.

While we expect that, at the level of the decay rate analysis, this approach of treating inflationary plateaus as additional vacua is equivalent to the previous discussion, it is not obvious how this goes together with the detailed balance assumption in Sect.~\ref{nvi}. This issue has been studied in the literature for Hawking-Moss (HM) \cite{Hawking:1981fz} transitions between adjacent vacua. In this case, detailed balance is obeyed \cite{Espinosa:2021tgx} and hence our generalization of Shnirelman's theorem in the form \eqref{Shnirelman_extended} might apply.
However, the real case of interest are slow-roll rather than HM-plateaus and we have to leave it to future work to determine whether an analogous property holds in this situation.

Ultimately, the clear goal for the future is to develop our brief discussion in Sects.~\ref{sect_toy_landscape} and \ref{sect_anthropic_prediction} to the level needed for deriving statements about anthropic observers in the string landscape.

\subsection{Related approaches}
Our construction has obvious similarities with \cite{nomura2011physical,nomura2012static}, \cite{Garriga:2005av,garriga2013watchers} and \cite{MersiniHoughton:2005im,Podolsky:2007vg}. It is also in part inspired by \cite{Hartle:2016tpo}. In the course of developing our perspective we have, of course, already commented on some of these results. Nevertheless, now that our picture is complete, it may be worthwhile providing a more explicit summary of similarities and differences.

The measure proposal of \cite{nomura2011physical, nomura2012static} is also based on finding the state of the multiverse living in a Hilbert space which takes a form similar to \eqref{Hilbert_space_structure}. However, in these references a key role is played by an `abstract observer' moving through the multiverse with the state of the universe describing the regions accessible to the `observer.' By contrast, our approach is observer-independent, as emphasized before. While \cite{nomura2011physical} assumes Schr\"odinger evolution with a global time variable, \cite{nomura2012static} refers to the WDW equation. But the selection conditions for the state of the multiverse are different: In our approach, $\Psi$ is a solution to the inhomogeneous equation~\eqref{iwdw} and it obeys appropriate boundary conditions, such that the state of the universe presumably is uniquely determined.
In \cite{nomura2012static}, $\Psi$ must solve the conventional, homogeneous WDW equation and be normalizable. If the WDW operator has multiple normalizable zero modes, there is a residual ambiguity in determining $\Psi$. This difference is related to the fact that the analysis in \cite{nomura2012static} does not consider the creation of dS vacua out of nothing. Such a possibility is discussed in \cite{nomura2011physical}, but the model built in this reference is not based on the WDW equation.
Crucially, in our analysis we are able to make contact between the more abstract WDW or Hilbert space approach and the semiclassical theory in the form of the simple rate equation \eqref{defining_equation}. This should provide a useful starting point for concrete applications.

Moreover, the proposal of \cite{nomura2012static} relies on special phenomena that were called `inverse bubbles.' These are pocket universes with contracting rather than expanding bubble walls. Such `inverse bubbles' are not well understood semiclassically and one of our main points is the proposal of a stationary solution with terminal vacua which nevertheless does not have to rely on such exotic objects.

In \cite{Garriga:2005av,garriga2013watchers}, rate equations similar to \eqref{defining_equation} appear. The analysis in these references is, however, based on the semiclassical perspective on the multiverse rather than quantum cosmology and the WDW equation. Hence, as discussed in more detail below \eqref{defining_equation}, the concepts leading to the resulting equation are different in our case.

Reference \cite{Hartle:2016tpo} also combines the creation of universes out of nothing with tunneling transitions in the multiverse to address the measure problem. We are at the moment unable to provide a detailed comparison to our approach since the analysis in \cite{Hartle:2016tpo} is implemented only in a toy model and it is not clear to us how it can be generalized. One aspect that we are struggling with in that approach is to understand how the small Hartle-Hawking 3-spheres are to be included in the very large 3-sphere of the late-time multiverse.

Simultaneous to our paper, reference \cite{Khoury:2022ish} was uploaded to the arxiv. It discusses the measure problem using Bayesian reasoning as the main ingredient. Although the underlying concepts are rather different compared to the local WDW approach, the resulting predictions appear to be similar.

Finally, also references \cite{MersiniHoughton:2005im} and \cite{Podolsky:2007vg} have starting points close to our analysis: the WDW equation and a local quantum-mechanical model respectively. But, as we already discussed in some detail at the end of Sect.~\ref{pwl}, the implementation is rather different. Let us only recall here that, as a key distinction, both references find Anderson localization in the landscape. It is then expected that only vacua around specific centers are populated. By contrast, since in our approach dS vacua can be created out of nothing, the entire landscape is constantly populated such that the probability distribution of vacua is not localized.

\section{Conclusions}
We have constructed a quantum mechanical model of the multiverse based on the Wheeler-DeWitt approach and the cosmological central dogma. 
The result is an inhomogeneous version of the WDW equation, the solution $\Psi$ of which lives in the Hilbert space
\begin{align}
		\mathcal{H}=\left(\bigoplus_{\,i\,\in\, {\it dS}} \mathcal{H}_{i}\right) \bigoplus\left(\bigoplus_{\,y\,\in\, \it Terminals}\mathcal{H}_{y}\right)\,,\label{Hilbert_space_structure}
\end{align}
with $\mbox{dim}(\mathcal{H}_i)=\exp(S_i)$ and $\mbox{dim}(\mathcal{H}_y)=\infty$.
Predictions for measurements are made by projecting $\Psi$ on the subspace of interest and subsequently taking its norm.
Each Hilbert space $\mathcal{H}_i$ in \eqref{Hilbert_space_structure} describes a static patch of dS vacuum $i$, such that the size of the expanding $3$-sphere of global dS space is irrelevant. 
Hence, the resulting measure for the multiverse can be called `local Wheeler-DeWitt measure.'
Motivated by semiclassical expectations, we allow for the creation of dS vacua out of nothing according to the no-boundary \cite{hartle1983wave} or tunneling \cite{Linde:1983mx,Vilenkin:1984wp,Vilenkin:1986cy} proposal, and we do not allow for transitions from terminal vacua back to the dS part of the Hilbert space. This forces $\Psi$ to obey specific boundary conditions which imply that the underlying fundamental equation is, as already mentioned, inhomogeneous. The inhomogeneous term encodes the fact that the probability current has sources in the dS-like part of the Hilbert space, cf.~\eqref{iwdw}. In addition, the current runs off to infinity in the terminal part of ${\cal H}$.

We note that the WDW approach to landscape dynamics also changes the perspective on the `everything happens' slogan and the spread of the wave function through the landscape. Indeed, in the stationary solution dictated by the inclusion of terminal vacua and the presence of sources, the probability current is `eternally flowing' for consistency, and the Universe does not require additional resources to find the anthropically viable vacua. The `static-patch, multiple vacua' point of view is both observer-independent and consistent with the cosmological version of the `central dogma.'

Since the microscopic Hamiltonian is unknown, it is a priori not obvious how to derive predictions from our model of the multiverse. We propose to achieve this by relating the quantum mechanical probability current to semiclassical decay rates, leading to a simple equation for the probability distribution of vacua -- eq.~\eqref{defining_equation}. This can be taken as the starting point for anthropic predictions,  such as the most likely vacuum for anthropic observers. In Sect.~\ref{sect_toy_landscape}, we have shown that the distribution of vacua appears to depend strongly on the chosen mechanism for creating dS vacua out of nothing. It would hence be crucial to determine on the theory side whether the Hartle-Hawking or the Linde/Vilenkin creation amplitude for dS vacua is the right one to be used in this context. Alternatively, it may become possible to make this decision on the basis of consistency with observations, given of course that our multiverse model is correct.

To summarize, the local WDW measure is a new proposal for making predictions in the multiverse. It has a clear and simple conceptual basis in quantum-cosmology. We expect that, after further reflection and discourse, the proposal will be refined and more details will emerge. Crucially, under certain assumptions to be scrutinized in the future, this proposal leads to explicitly calculable probabilities. Its validity can hence be established by comparing anthropic predictions to observation.

\subsection*{Acknowledgements}
This work was supported by Deutsche Forschungsgemeinschaft (DFG, German Research Foundation) under Germany’s Excellence Strategy EXC 2181/1 - 390900948 (the Heidelberg STRUCTURES Excellence Cluster). 
A.H. gratefully acknowledges very useful discussions and unpublished work on the measure problem with Alexander Westphal.
We thank Justin Khoury and Sam S. C. Wong for helpful communications.

\bibliographystyle{utphys}
\bibliography{Predictions_toy_model_dS}
\end{document}